\documentclass[twocolumn]{article}
\usepackage{cite}
\usepackage[pdftex]{graphicx}
\usepackage[]{subfig}
\usepackage{ragged2e}
\usepackage{array}
\newcolumntype{P}[1]{>{\centering\arraybackslash}p{#1}}
\newcolumntype{M}[1]{>{\centering\arraybackslash}m{#1}}
\usepackage{multirow}
\usepackage[normalem]{ulem}
\useunder{\uline}{\ul}{}
\usepackage[fleqn]{amsmath}
\usepackage{newtxtext}
\usepackage[varg]{newtxmath}
\usepackage{algorithm,algpseudocode}
\usepackage{multirow}
\usepackage{array}
\usepackage[table]{xcolor}
\usepackage{xspace}
\usepackage{authblk}

\newcommand{\ssimVP}{{SSIM}\xspace}
\newcommand{\mssimVP}{{MS-SSIM}\xspace}
\newcommand{\mseVP}{{MSE}\xspace}
\newcommand{\psnrVP}{{VPSNR}\xspace}
\newcommand{\uqiVP}{{UQI}\xspace} 
\newcommand{\iwSsimVP}{{IW-SSIM}\xspace} 
 
\newcommand{\iwPnsrVP}{{IW-PSNR}\xspace} 
\newcommand{\fsimVP}{{FSIM}\xspace} 
\newcommand{\fsimCVP}{{FSIMc}\xspace} 
\newcommand{\rfsimVP}{{RFSIM}\xspace} 
\newcommand{\vifpVP}{{VIFp}\xspace} 
\newcommand{\vipVP}{{VIF}\xspace} 
\newcommand{\srSimVP}{{SR-SIM}\xspace} 
\newcommand{\nqmVP}{{NQM}\xspace} 
\newcommand{\fwqiVP}{{FWQI}\xspace} 
\newcommand{\fSsimVP}{{F-SSIM}\xspace} 
\newcommand{\fwSnrVP}{{FWSNR}\xspace} 
\newcommand{\wsnrVP}{{WSNR}\xspace} 
\newcommand{\fpsnrVP}{{FPSNR}\xspace} 

\newcommand{\psnr}{PSNR\xspace}
\newcommand{\wpsnr}{\textit{ZWF}\xspace}

\usepackage{amsmath}
\usepackage{siunitx}
\makeatletter
\providecommand\add@text{}
\newcommand\tagaddtext[1]{%
	\gdef\add@text{#1\gdef\add@text{}}}%
\renewcommand\tagform@[1]{%
	\maketag@@@{\llap{\add@text\quad}(\ignorespaces#1\unskip\@@italiccorr)}%
}
\makeatother

\title{Impacts of Retina-related Zones on Quality Perception of Omnidirectional Image}
\author[1]{Huyen~T.~T.~Tran}
\author[1]{Duc V. Nguyen}
\author[2]{Nam~Pham~Ngoc}
\author[3]{Trang~H.~Hoang}
\author[3]{Truong Thu Huong}
\author[1]{Truong~Cong~Thang}
\affil[1]{The University of Aizu, Aizuwakamatsu, Japan}
\affil[2]{Vin-University project, Vietnam}
\affil[3]{Hanoi University of Science and Technology, Hanoi, Vietnam}
\date{}                     
\begin{document}	

	\maketitle
	
	\begin{abstract}
		Virtual Reality (VR), which brings immersive experiences to viewers, has been gaining popularity in recent years. A key feature in VR systems is the use of omnidirectional content, which provides 360-degree views of scenes. In this work, we study the human quality perception of omnidirectional images, focusing on different zones surrounding the foveation point. For that purpose, an extensive subjective experiment is carried out to assess the perceptual quality of omnidirectional images with non-uniform quality. Through experimental results, the impacts of different zones are analyzed. Moreover, nineteen objective quality metrics, including foveal quality metrics, are evaluated using our database. It is quantitatively shown that the zones corresponding to the fovea and parafovea of human eyes are extremely important for quality perception, while the impacts of the other zones corresponding to the perifovea and periphery are small. Besides, the investigated metrics are found to be not effective enough to reflect the quality perceived by viewers.
	\end{abstract}

	\section{Introduction}\label{SectIntro}
	In order to bring immersive experiences to viewers, virtual reality (VR) systems employ omnidirectional content which contains 360-degree views of scenes. Unlike traditional content displayed using a flat screen, omnidirectional content is usually consumed using Head Mounted Displays (HMDs). Also, only a small part of the full content (called \textit{viewport}) corresponding to the current viewing direction is actually seen by the viewer at a moment~\cite{DucISM}.  
	
	Because omnidirectional (or 360-degree) content has very high bitrate, a key challenge in omnidirectional content delivery is how to optimize system resources while still ensuring satisfactory user experience. For that, many encoding and delivery solutions have been proposed in the literature, where the (estimated) viewport is provided with high quality and the remaining part with low quality~\cite{JacobAdapt,ozcinar2017viewport,Duc_Jetcas}. Moreover, in VR systems, foveated imaging, which decreases quality of zones far from the viewer's foveation point~\cite{3DFoveated,NVIDIA2017latency}, can be used to further reduce resource consumption~\cite{WangTIP360,NVIDIA2017latency}. However, the estimated viewing direction could be very different from the actual one when the system delay is large~\cite{DucTron}. Even the viewer may suddenly turn to look at the back.  In these cases, the actual viewport may have low quality in the central part and high quality in the periphery. In other words, the central part may have higher quality (called scenario $S\#1$) or lower quality (called scenario $S\#2$) than the periphery, both resulting in omnidirectional content with non-uniform quality.  
	
	It is well-known that human visual acuity is spatially variable~\cite{LetterVisualAccutiy,LeeFilter}. In particular, when a person gazes at a point, called \textit{foveation point}, a zone closer to this point is perceived to be sharper than the others. This means that the human eyes have a higher sensitivity to distortions in the central than in the periphery. Hence, the understanding of the impacts of different zones on the perceptual quality is obviously of indispensable necessity in the context of omnidirectional content.   
	
	In the literature, there are only a few existing studies on subjective quality assessments of images/videos with non-uniform quality~\cite{Lausane2011FoveatedCodingEvaluation,ChihFoveated2017,WangTIP360}. However, most of these studies are devoted to traditional content~\cite{Lausane2011FoveatedCodingEvaluation,ChihFoveated2017}. In~\cite{Lausane2011FoveatedCodingEvaluation}, each image is divided into four zones of equal widths. The quality of these zones gradually decreases with a fixed step size. It is found that, when the step size is small, the  difference of perceptual quality between the non-uniform and uniform videos is insignificant. In addition, the maximum value of the step size without causing significant quality differences depends on content characteristics. In~\cite{ChihFoveated2017}, each image is divided into three zones, which are foveal, blending, and peripheral zones. Through experimental results, the finding is that participants barely notice quality decreases at the peripheral zones of the eccentricity larger than 7.5~degrees. Also, an evaluation of four subjective assessment methods is presented. It is indicated that the Absolute Category Rating (ACR) method is the best method to evaluate the subjective quality of non-uniform images.            
	
	In the literature, there have been some studies on subjective quality assessments of omnidirectional content ~\cite{JiaoTong_VRDB,MaixuVR2019,Kaist2019VR,ViewDuration20s}. In these studies, various distortion types such as compression and Gaussian blur are considered. However, the distortions are distributed uniformly in~\cite{JiaoTong_VRDB,MaixuVR2019,Kaist2019VR,ViewDuration20s}. The work in~\cite{WangTIP360} is the only previous study on omnidirectional content with non-uniform quality. In~\cite{WangTIP360}, the authors focus on answering the question of how to spatially reduce image quality without causing impacts on user perception. For that purpose, they propose to divide an omnidirectional image into three areas according to three regions of the human retina, namely the macula, the near periphery, and the far periphery. The image quality corresponding to each region is decreased step by step until participants notice a perceptual difference. The encoding parameters obtained just before that point are modeled and then used as a guide for spatially reducing image quality without perceptual loss. It is shown that this approach could save loading time about 90\% comparing to a conventional approach using uniform quality.
	
	Over several decades, a large number of objective quality metrics have been proposed~\cite{ssim,msssim,uqi,LeeFWSNR,fssim}. Some of these metrics take into account the foveation feature, hereafter referred to as \textit{foveal quality metrics}~\cite{LeeFWSNR,fssim}. However, all these metrics are specific to traditional content. There has been no existing foveal quality metric for omnidirectional content so far. 
	
	In our previous study~\cite{Huyen_VRMetrics}, a comparison between eight state-of-the-art quality metrics has been conducted. Experimental results show that \psnr turns out to be the most effective metric for quality assessment of omnidirectional videos. However, it is worth to note that stimuli used in that study have uniform quality. As shown later in this paper, \psnr is actually not effective when the quality is spatially variable. To the best of our knowledge, no extensive evaluation of objective quality metrics for omnidirectional images with non-uniform quality has been conducted in the literature.  
	
	In this study, our purposes related to user perception of omnidirectional content in VR systems include: 
	\begin{itemize}
		\item Subjective study on the impacts of retina-related zones on quality perception of omnidirectional images. 
		\item Performance evaluation of existing objective quality metrics, especially foveal quality metrics, for omnidirectional images having non-uniform quality. 
	\end{itemize}
	
	To that end, our major contributions are as follows. First, we present a detailed description of a VR viewing geometry and the human retina. This description helps in designing subjective experiments and in calculating parameters used in foveal quality metrics.     
	Second, we carry out an extensive subjective experiment with 256 stimuli of non-uniform quality. The quality zones of the stimuli are designed based on five regions of the human retina. 
	Third, using a simple zone-weighted formulation, we quantify the impacts of different zones on the perceptual quality. It is quantitatively found that the zones corresponding to the fovea and parafovea of the human retina are extremely important for quality perception. Also, the impacts of zones are strongly affected by content characteristics.
	Fourth, we evaluate the correlation of nineteen objective quality metrics against subjective scores. Experimental results indicate that these metrics, even the foveal ones, are not very effective when the viewport quality is spatially variable.   
	
	The remainder of the paper is organized as follows. A description of a VR viewing geometry and the human retina is presented in Sect.~\ref{SectOverview}. Sect.~\ref{SectExperiment} presents the details of the subjective experiment. The analysis of perceptual behaviors using the experimental results is provided in Sect.~\ref{SectAnalysis}. Then, an evaluation of quality metrics is presented in Sect.~\ref{SectEvaluation}. Section~\ref{SectConclusion} concludes the paper and provides an outlook on future work.

	\section{Overview}\label{SectOverview}
	\textcolor{black}{In this section, the viewing geometry in VR systems is first presented. Then, the regions in human retina are described.}
	\subsection{Viewing Geometry in VR Systems}\label{subSectViewingGeometry}
\begin{figure}[t]
		\centering
		\includegraphics[width=\columnwidth]{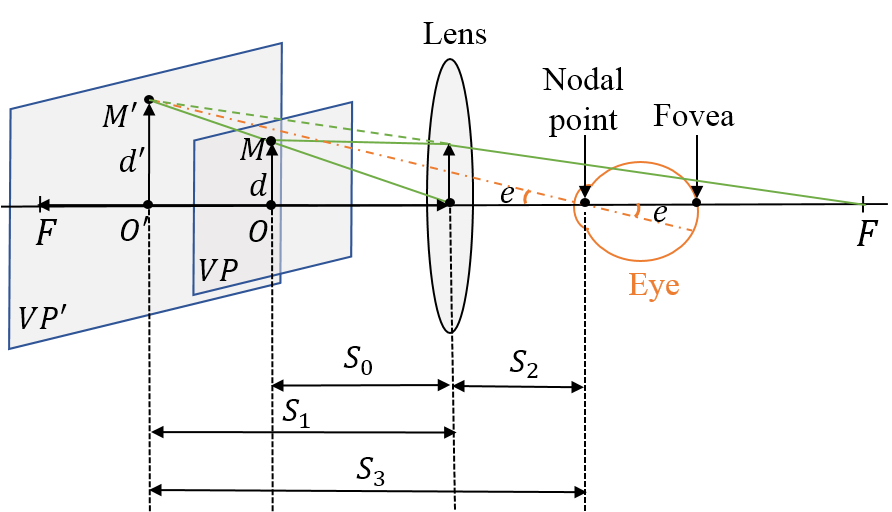}
		\caption{Typical viewing geometry in VR systems}
		\label{figViewingGeometry}
	\end{figure}

	Fig.~\ref{figViewingGeometry} illustrates a typical viewing geometry in VR systems. Assume that $\textit{VP}$ is the displayed viewport, the lens in the HMD produces a virtual viewport $\textit{VP}'$ that is further formed on the retina in the human eyes. Eccentricity $e$ (degrees) is used to measure the angular distance from the central gaze direction to any point in the virtual viewport $\textit{VP}'$. 
	
	Let $F$ (units of length) be the focal length of the lens. $S_0$, $S_1$, and $S_2$ (units of length) respectively denote the distances from the lens to the displayed viewport $\textit{VP}$, the virtual viewport $\textit{VP}'$, and the eye. Based on lens equations, the distance from the lens to the virtual viewport $ S_1 $ is computed by 
	\begin{equation}
	S_1= S_0 \times \frac{F}{F-S_0}~~~~~~\tagaddtext{[units of length].}        
	\end{equation}
	Then, the distance from the eye to the virtual viewport is calculated by
	\begin{equation}
	S_3=S_1+S_2 \tagaddtext{[units of length].}        
	\end{equation}

	Let $W_p{\times}H_p$ (pixels) and $W_l{\times}H_l$ (units of length) respectively be the width and height of the displayed viewport $\textit{VP}$ in pixels and units of length. The width of the virtual viewport $\textit{VP}'$ (in pixels and units of length) is given by the following equations. 
	\begin{equation}
	W_p'=W_p \tagaddtext{[pixels].}
	\end{equation}
	\begin{equation}
	W_l'=W_l\times \frac{F}{F-S_0}~~~~~~\tagaddtext{[units of length].}
	\end{equation}
	Also, the height of the virtual viewport $\textit{VP}'$ is calculated by
	\begin{equation}
	H_p'=H_p \tagaddtext{[pixels]}
	\end{equation} 
	and 
	\begin{equation}
	H_l'=H_l\times \frac{F}{F-S_0}~~~~~~\tagaddtext{[units of length].}
	\end{equation}

	Assume that the foveation point is the center $O'=(x_{O'},y_{O'})$ in the virtual viewport $\textit{VP}'$. Point $O=(x_O,y_O)$ in the displayed viewport $\textit{VP}$ corresponding to point $O'$ is determined by 
	\begin{equation}
	x_O = x_{O'}	\tagaddtext{[pixels]}
	\end{equation}
	and 
	\begin{equation}
	y_O = y_{O'}	\tagaddtext{[pixels].}
	\end{equation}
	
	Let $M$ be a point at the position of $(x_M,y_M)$  (pixels) in the displayed viewport $\textit{VP}$. The position of the virtual point $M'=(x_{M'},y_{M'})$ corresponding to point $M$ is  
	\begin{equation}
	x_{M'} = x_M	\tagaddtext{[pixels]}
	\end{equation}
	and 
	\begin{equation}
	y_{M'} = y_M \tagaddtext{[pixels].}
	\end{equation}

	The distance from pixel $M'$ to the foveation point $O'$ is 
	\begin{equation}
	\small
	\begin{aligned}
	& d'=\sqrt{\left(\frac{(x_{M'}-x_{O'})\times W_l'}{W_p'}\right)^2+\left(\frac{(y_{M'}-y_{O'})\times H_l'}{H_p'}\right)^2} \\
	&~~~~~~~~~~~~~~~~~~~~~~~~~~~~~~~~~~~~~~~~~~~~~~~~~~~\text{[units of length].}
	\end{aligned}
	\end{equation}

	The eccentricity $e$ of point $M'$ in the virtual viewport $\textit{VP}'$ is given by
	\begin{equation}
	e(x_{M'},y_{M'}) = tan^{-1}\left(\frac{d'}{S_3}\right) \tagaddtext{[degrees].}  
	\end{equation}   
	
	It should be noted that parameters of a point on the virtual viewport are what actually used in a foveal quality metric. Moreover, given the knowledge of the human visual system, points on the virtual viewport can be divided according to the regions of the retina.

	\subsection{Regions in Human Retina}\label{SectRetina}
	\begin{figure}[t]
		\centering
		\includegraphics[width=0.8\columnwidth]{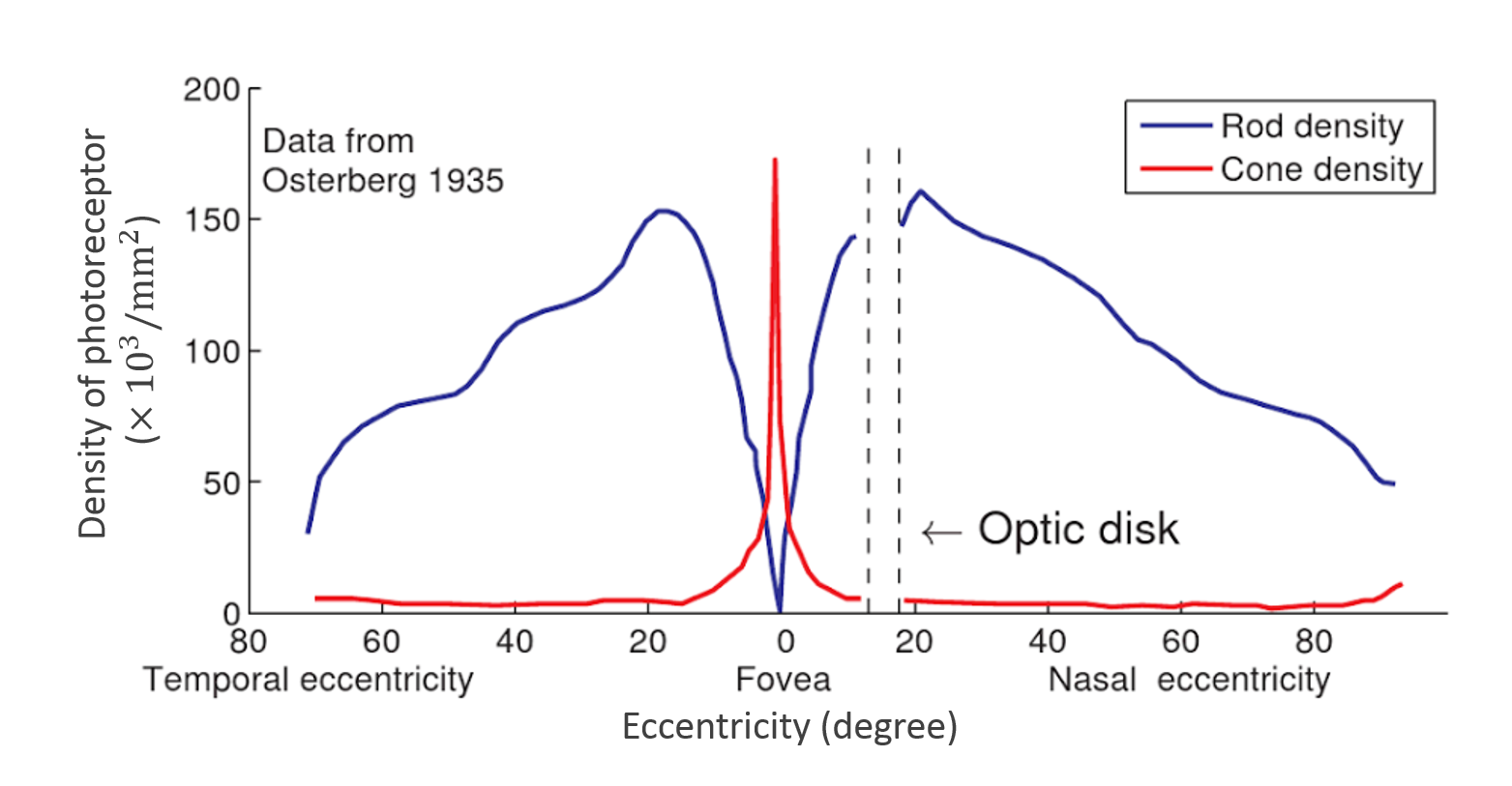}%
		\caption{Density of photoreceptors in the retinal~\cite{HVS_Density}}
		\label{figDensity}
	\end{figure}

	\begin{figure}[t]
		\centering
		\includegraphics[width=0.8\columnwidth]{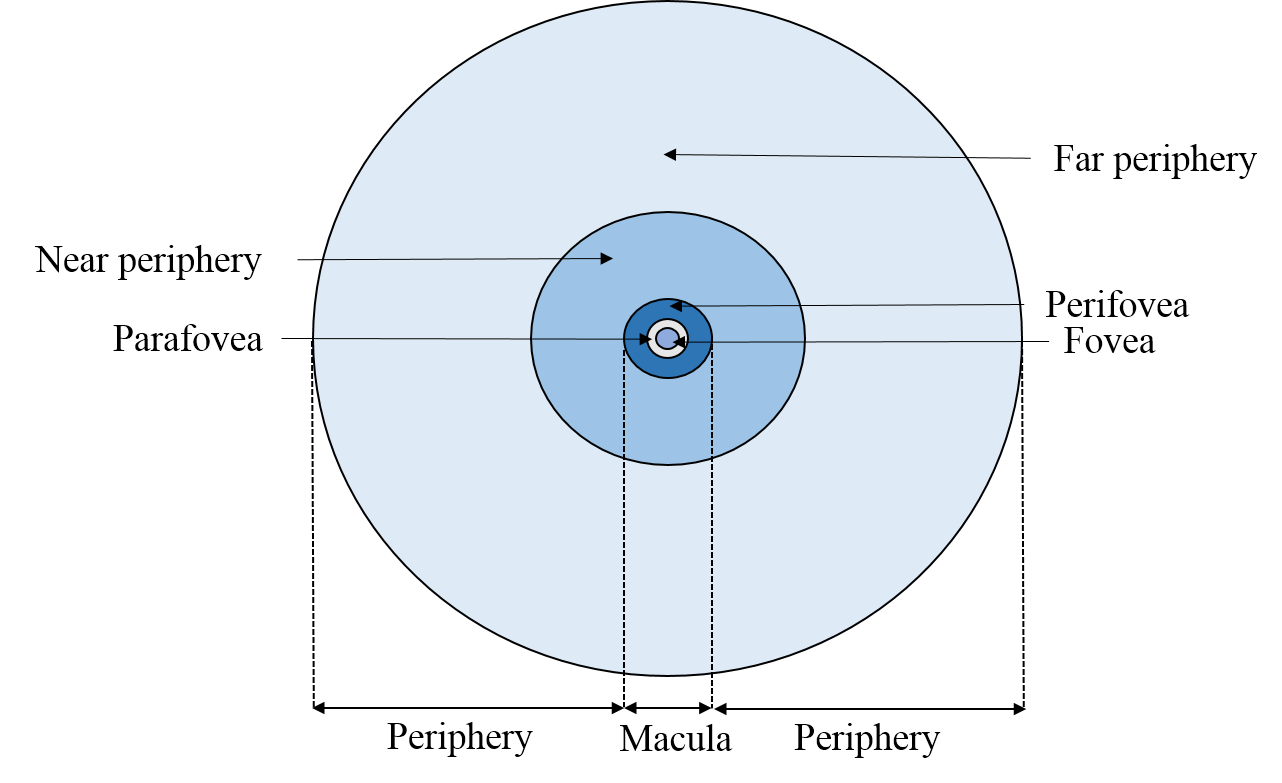}%
		\caption{Five regions of the retina}
		\label{figRegionsFoveal}
	\end{figure}
	
	In the human retina, there are two types of photoreceptors, namely rods and cones, each plays an important role in human visual system. In particular, cones function most effectively in relatively bright light and are responsible for color vision and visual acuity. Meanwhile, rods have higher sensitivities to light, and  thus they function mainly in dim light.
	
	Fig.~\ref{figDensity} shows the density of photoreceptors in the human retina. It can be seen that most cones are concentrated at the center of the retina, whereas rods are located away from the center. Visual information from photoreceptors are then collected by the so-called ganglion cells. The optic disk is where axons from ganglion cells exit the retina and convey visual information to the brain.

	\begin{figure*}[t]
		\centering\centering
		\subfloat[I1]{\includegraphics[width=0.22\textwidth,height=0.11\textwidth, keepaspectratio=false]{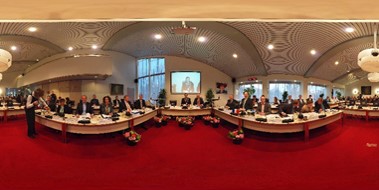}%
			\label{figI1_o}}
		\hfil
		\subfloat[I2]{\includegraphics[width=0.22\textwidth,height=0.11\textwidth, keepaspectratio=false]{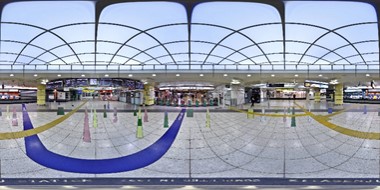}%
			\label{figI2_o}}
		\hfil
		\subfloat[I3]{\includegraphics[width=0.22\textwidth,height=0.11\textwidth, keepaspectratio=false]{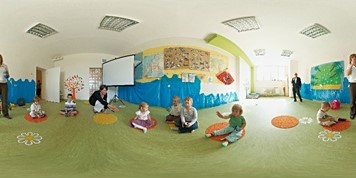}%
			\label{figI3_o}}
		\hfil
		\subfloat[I4]{\includegraphics[width=0.22\textwidth,height=0.11\textwidth, keepaspectratio=false]{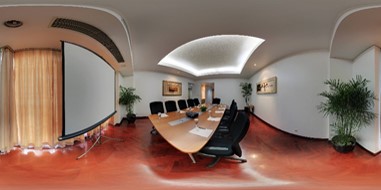}%
			\label{figI4_o}}
		\hfil \\
		\subfloat[I5]{\includegraphics[width=0.22\textwidth,height=0.11\textwidth, keepaspectratio=false]{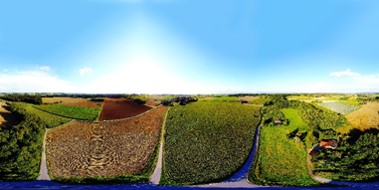}%
			\label{figI5_o}}
		\hfil
		\subfloat[I6]{\includegraphics[width=0.22\textwidth,height=0.11\textwidth, keepaspectratio=false]{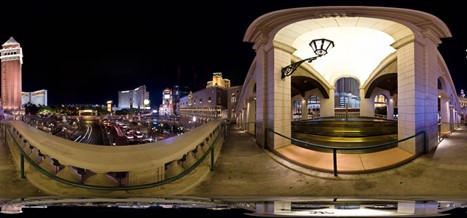}%
			\label{figI6_o}}
		\hfil
		\subfloat[I7]{\includegraphics[width=0.22\textwidth,height=0.11\textwidth, keepaspectratio=false]{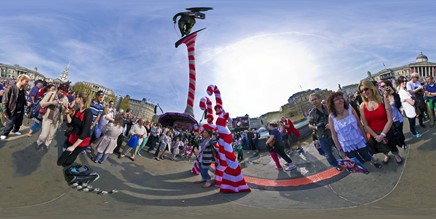}%
			\label{figI7_o}}
		\hfil
		\subfloat[I8]{\includegraphics[width=0.22\textwidth,height=0.11\textwidth, keepaspectratio=false]{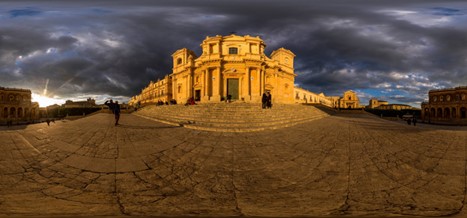}%
			\label{figI8_o}}
		\caption{Eight omnidirectional images used in our experiment}
		\label{figImage}
	\end{figure*}

	Based on the ganglion cell layer, the retina of human eyes can be divided into two main parts, namely macula and periphery~\cite{HVS_Regions}, as illustrated in Fig.~\ref{figRegionsFoveal}. In particular, the ganglion cell layer in the macula is several cells thick. Meanwhile, the periphery is only one ganglion cell thick. 
	The macula is further divided into three regions, called fovea, parafovea, and perifovea. The periphery is in turn divided into two regions, namely near periphery and far periphery~\cite{FoveaEye,HVS_Regions}. These five regions of the retina are briefly described below.  It is worth noting that there has been no standard definition of boundaries between these regions so far~\cite{ReviewRegionEyes}. In our research, the boundaries are determined based on~\cite{ReviewRegionEyes,RegionEyeVito,poppel1973light,jones2013peripheral}.

	The fovea is a small central region of the macula that represents 5~degrees of the central visual field or an eccentricity interval between 0~degree and 2.5~degrees. This region consists of densely packed cones. In addition, it has a layer of ganglion cells, which can be up to eight cells thick. Therefore, the fovea vision has the highest sensitivity to fine details. 
	
	The fovea is surrounded by the parafovea belt corresponding to an eccentricity interval between 2.5~degrees and 4~degrees. In the parafovea, rods are more numerous. Meanwhile, the thickness of the ganglion cell layer decreases from eight to four cells at its outer edge~\cite{FoveaEye}. 
	
	The region next to the parafovea is the perifovea with the corresponding eccentricity interval between 4~degrees and 9~degrees. In this region, the density of rods is higher than that of cones. The thickness of ganglion cell layer reduces to one cell at its peripheral edge~\cite{FoveaEye}. 
	
	In the periphery, the region corresponding to an eccentricity interval between 9~degrees and 30~degrees is the near periphery, and the rest is the far periphery. The dividing line corresponding to the eccentricity of 30~degrees is selected based on several features of visual performance. In particular, letter visual acuity decreases linearly with eccentricity from 0~degree to 30~degrees. For eccentricities larger than 30~degrees, the decrease is much steeper~\cite{LetterVisualAccutiy}.   
	
	Based on the above description of the viewing geometry and the retina, stimuli used in the following subjective experiment are designed so that the zones in the virtual viewports will correspond to the five regions of the retina. It is worth noting that, in this paper, we focus on the contributions (or weights) of different zones in the perceptual quality, rather than the quality-reducing trends as in~\cite{Lausane2011FoveatedCodingEvaluation,ChihFoveated2017,WangTIP360}.

	\begin{table}[t]
		\caption{Features of source images}
		\begin{center}
			\resizebox{\columnwidth}{!}{%
			\begin{tabular}{|M{0.5in}|m{4in}|}
				\hline
				\textbf{Image} &  \multicolumn{1}{c|}{\textbf{Description}}   \\ \hline
				I1      & indoor scene, large conference room, containing human faces \\ \hline
				I2      & indoor scene, train station in Japan, containing human faces \\ \hline
				I3      & indoor scene, small kindergarten classroom, containing human faces \\ \hline
				I4      & indoor scene, meeting room, without presence of human\\ \hline
				I5      & outdoor scene, natural landscape, daytime, without presence of human \\ \hline
				I6      & outdoor scene, balcony, nighttime, without presence of human \\ \hline
				I7      & outdoor scene, festival, daytime, containing human faces \\ \hline
				I8      & outdoor scene, outside of a cathedral, at sunset, containing human faces\\ \hline
			\end{tabular}
			}
		\end{center}
		\label{tabFeatureSourImage}
	\end{table}

	\begin{table}[t]
		\caption{Eccentricity intervals of zones}
		\centering
		\resizebox{\columnwidth}{!}{%
		\begin{tabular}{|M{1.5in}|c|c|c|c|c|}
			\hline
			Zone & $Z_1$ & $Z_2$ & $Z_3$& $Z_4$ & $Z_5$ \\ \hline
			Eccentricity interval (degrees) & [0,2.5) & [2.5,4) & [4,9) & [9,30) & [30,$+\infty$) \\ \hline
		\end{tabular}
		}
		\label{tabEcenIntervalZones}
	\end{table}
	
	\begin{figure*}[t]
		\centering\centering
		\subfloat[I1]{\includegraphics[width=0.12\textwidth]{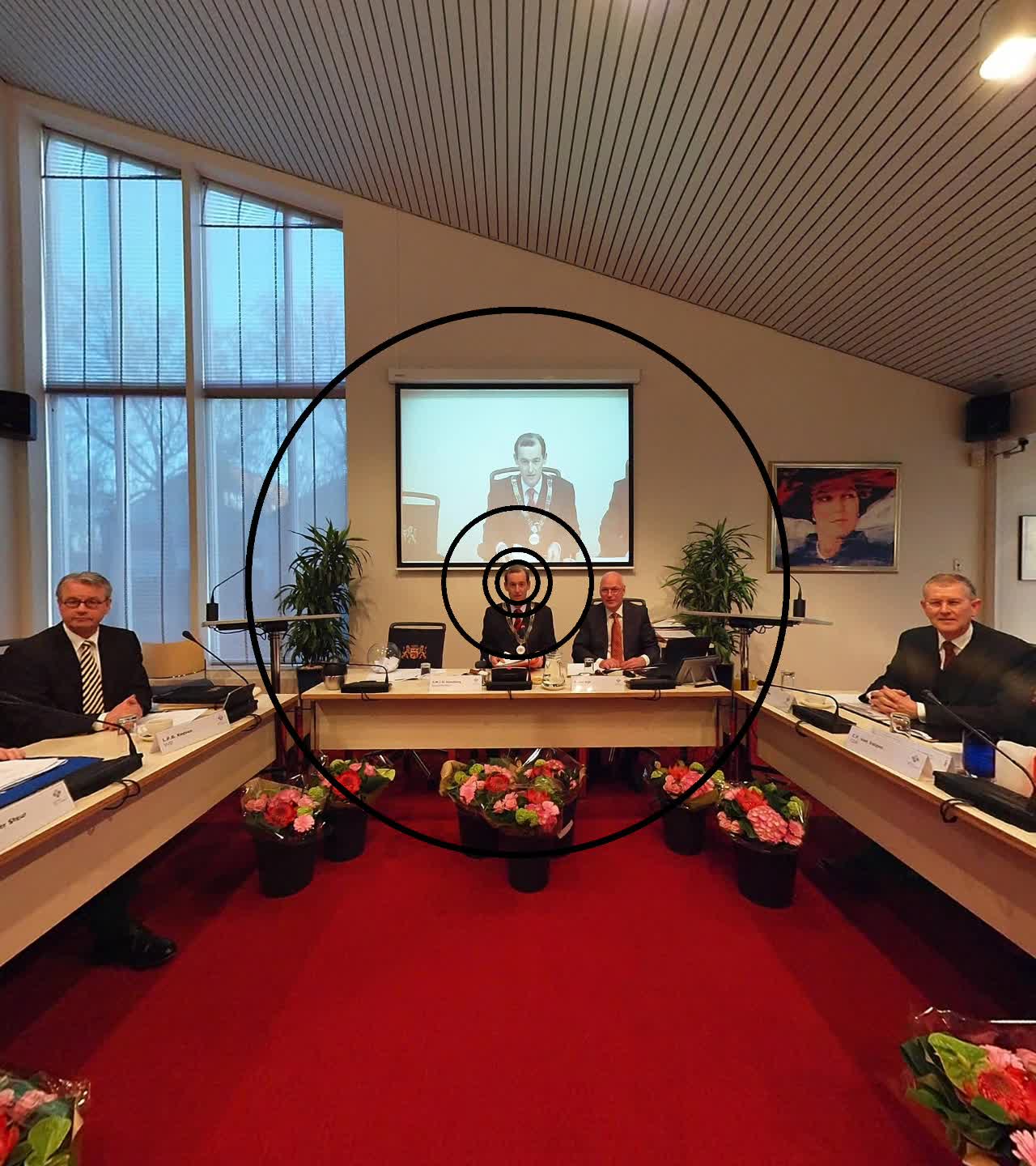}}
		\hfil
		\subfloat[I2]{\includegraphics[width=0.12\textwidth]{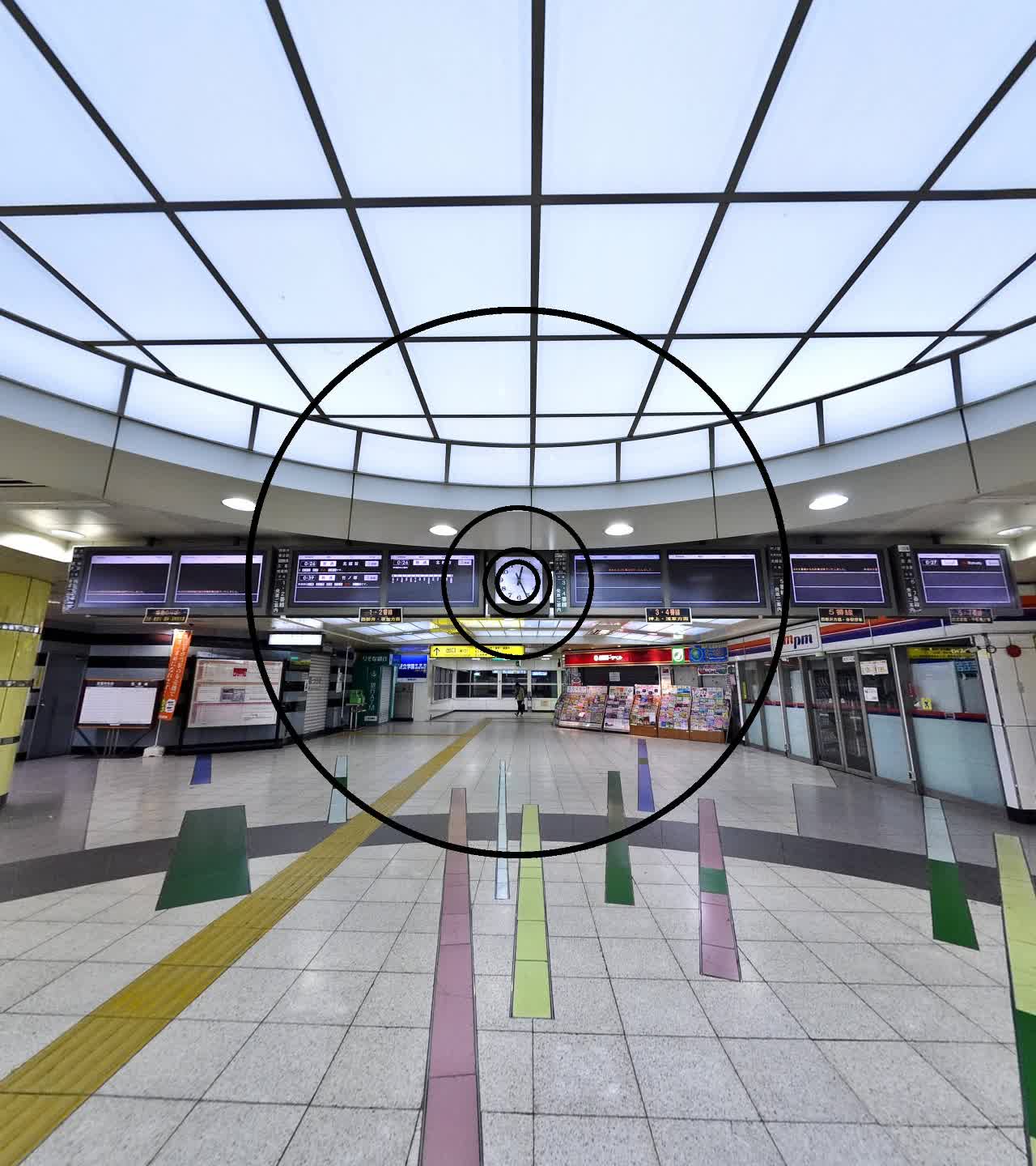}}
		\hfil
		\subfloat[I3]{\includegraphics[width=0.12\textwidth]{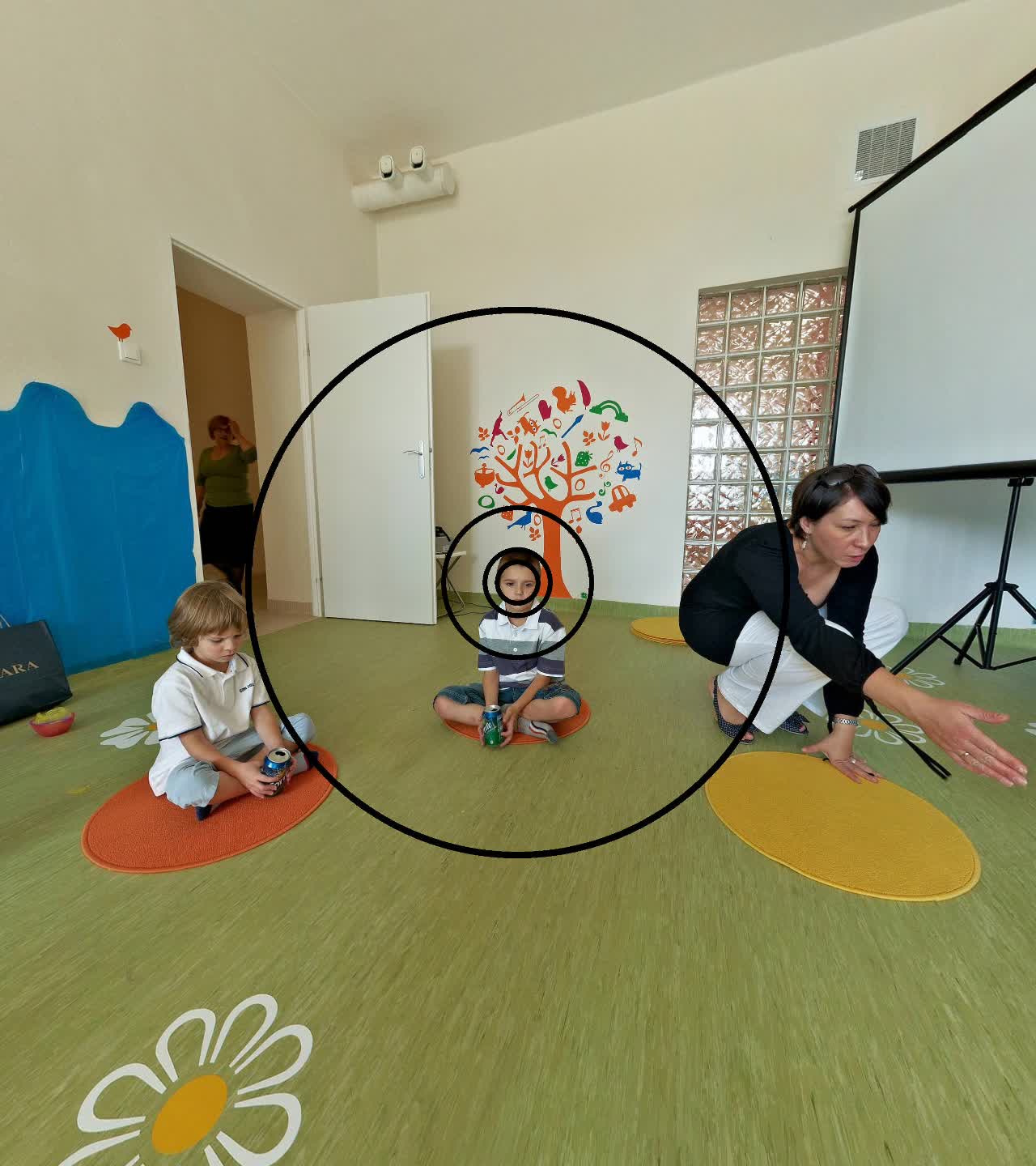}}
		\hfil
		\subfloat[I4]{\includegraphics[width=0.12\textwidth]{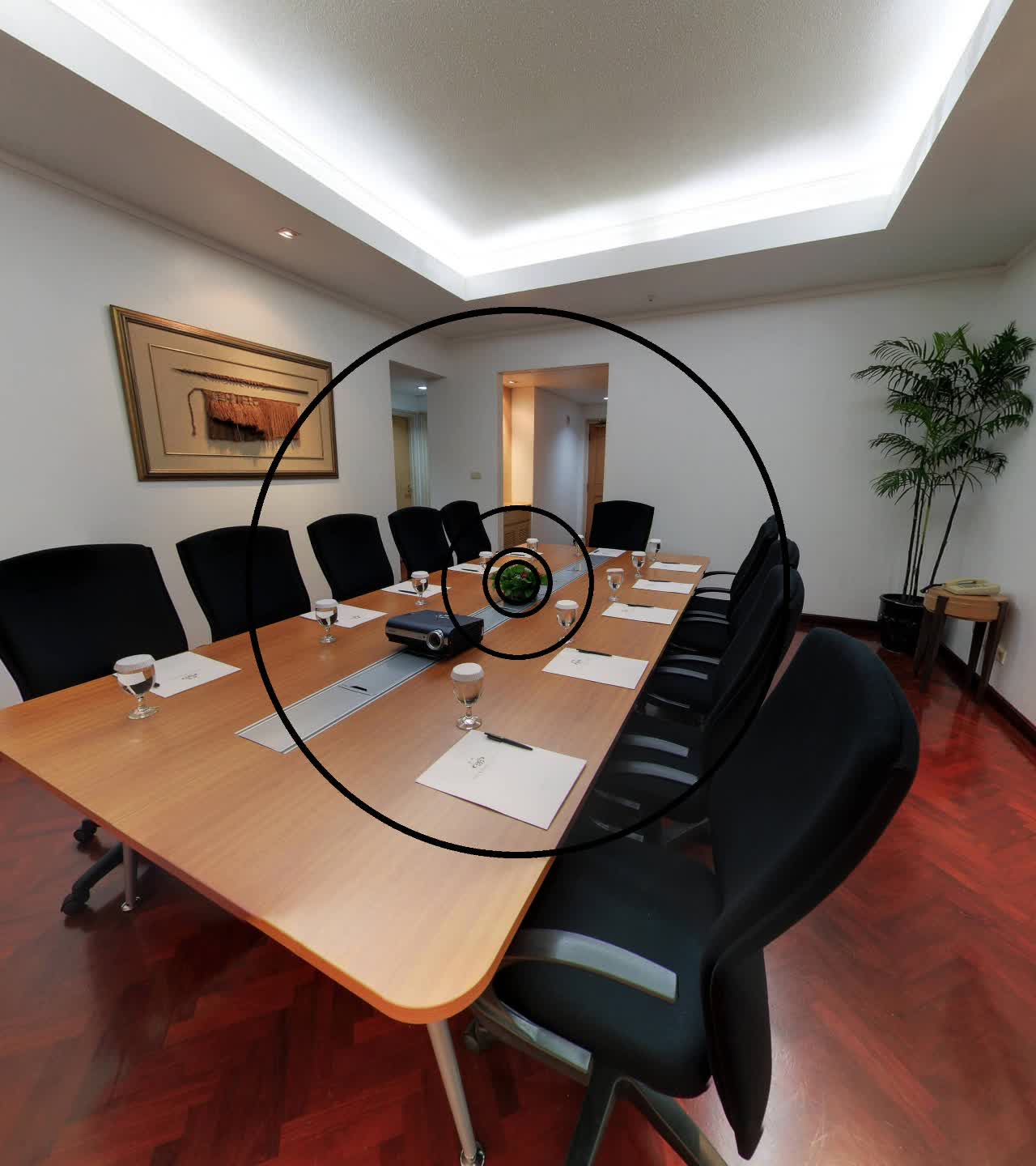}}
		\hfil
		\subfloat[I5]{\includegraphics[width=0.12\textwidth]{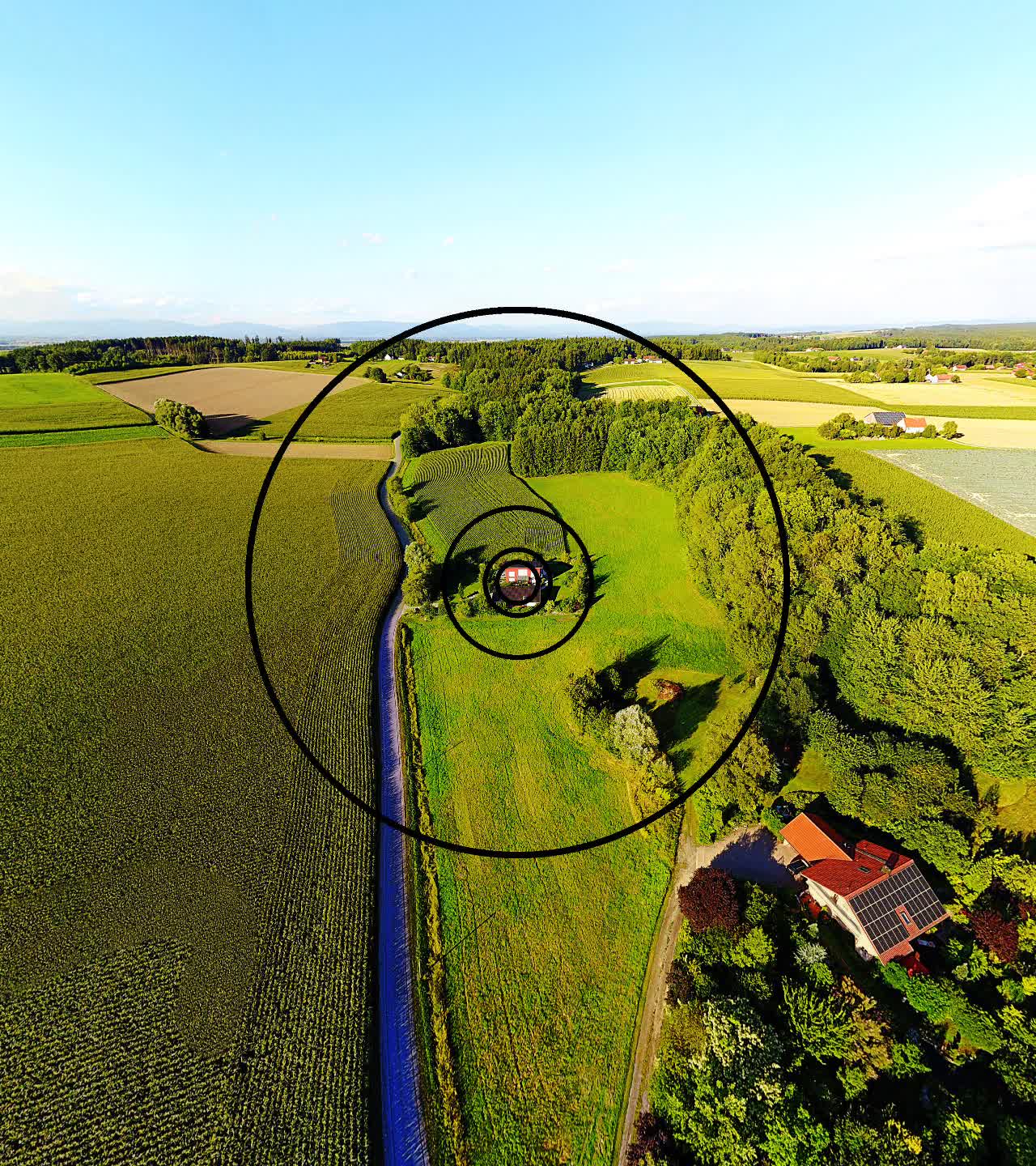}}
		\hfil
		\subfloat[I6]{\includegraphics[width=0.12\textwidth]{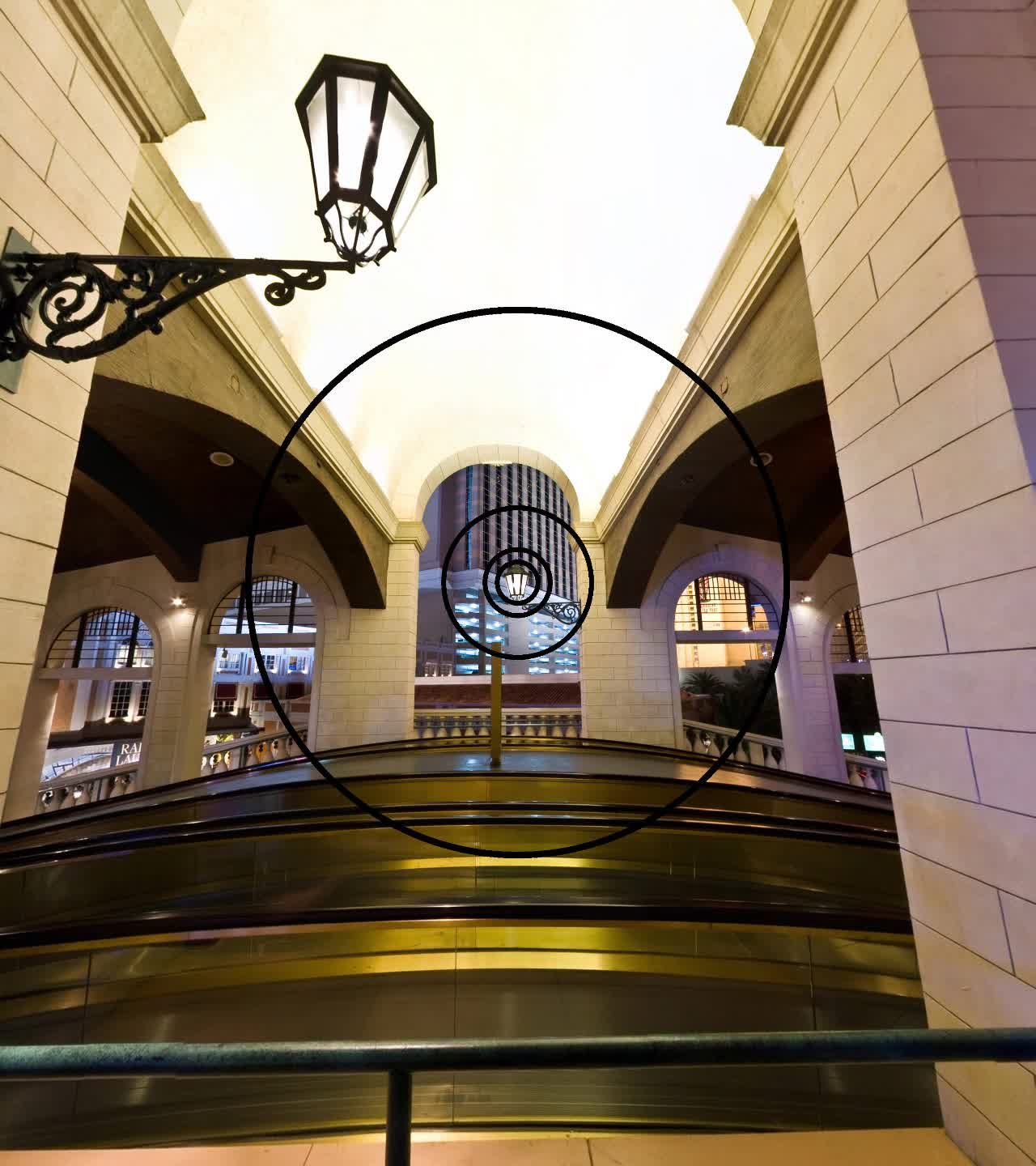}}
		\hfil
		\subfloat[I7]{\includegraphics[width=0.12\textwidth]{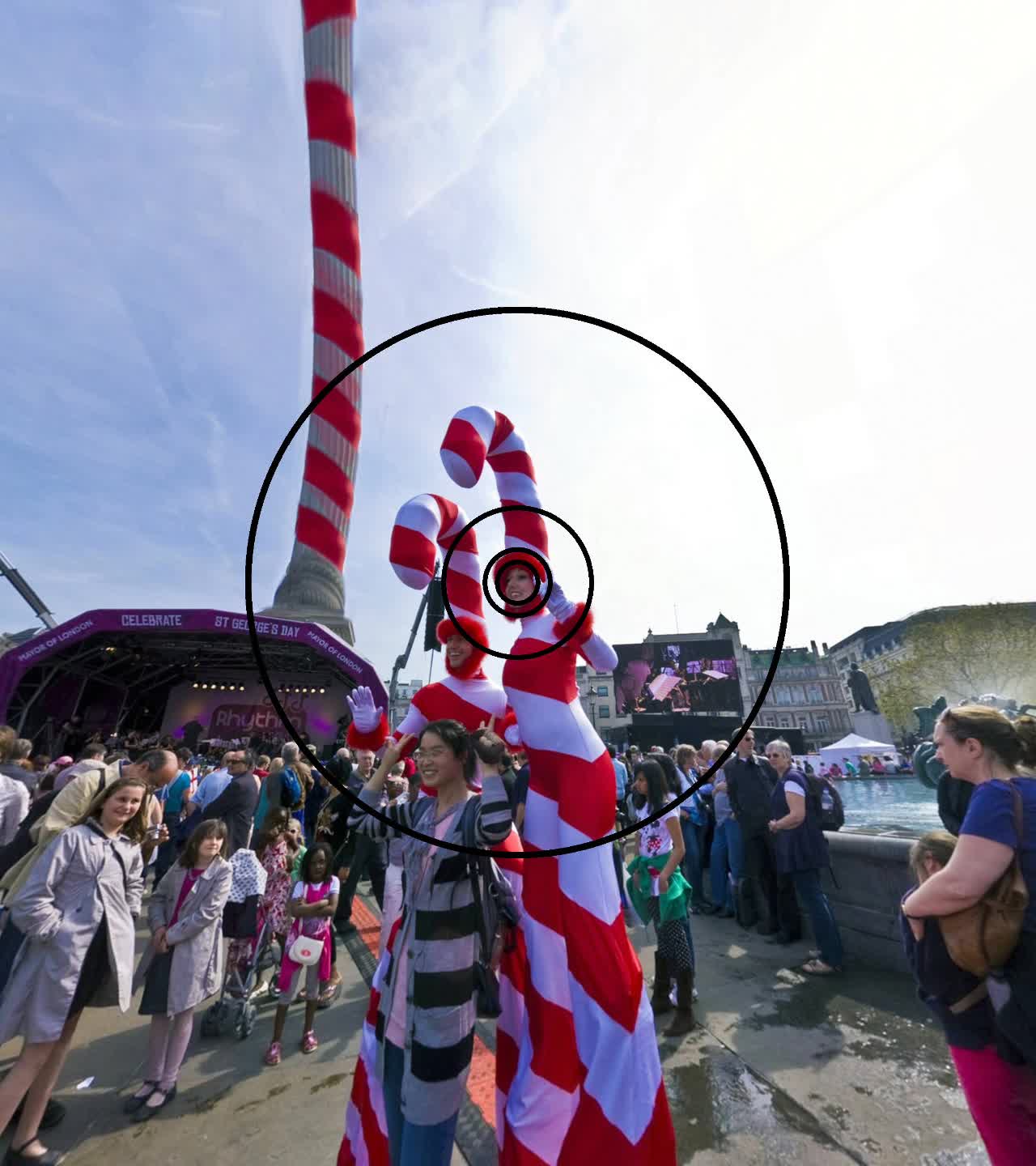}}
		\hfil
		\subfloat[I8]{\includegraphics[width=0.12\textwidth]{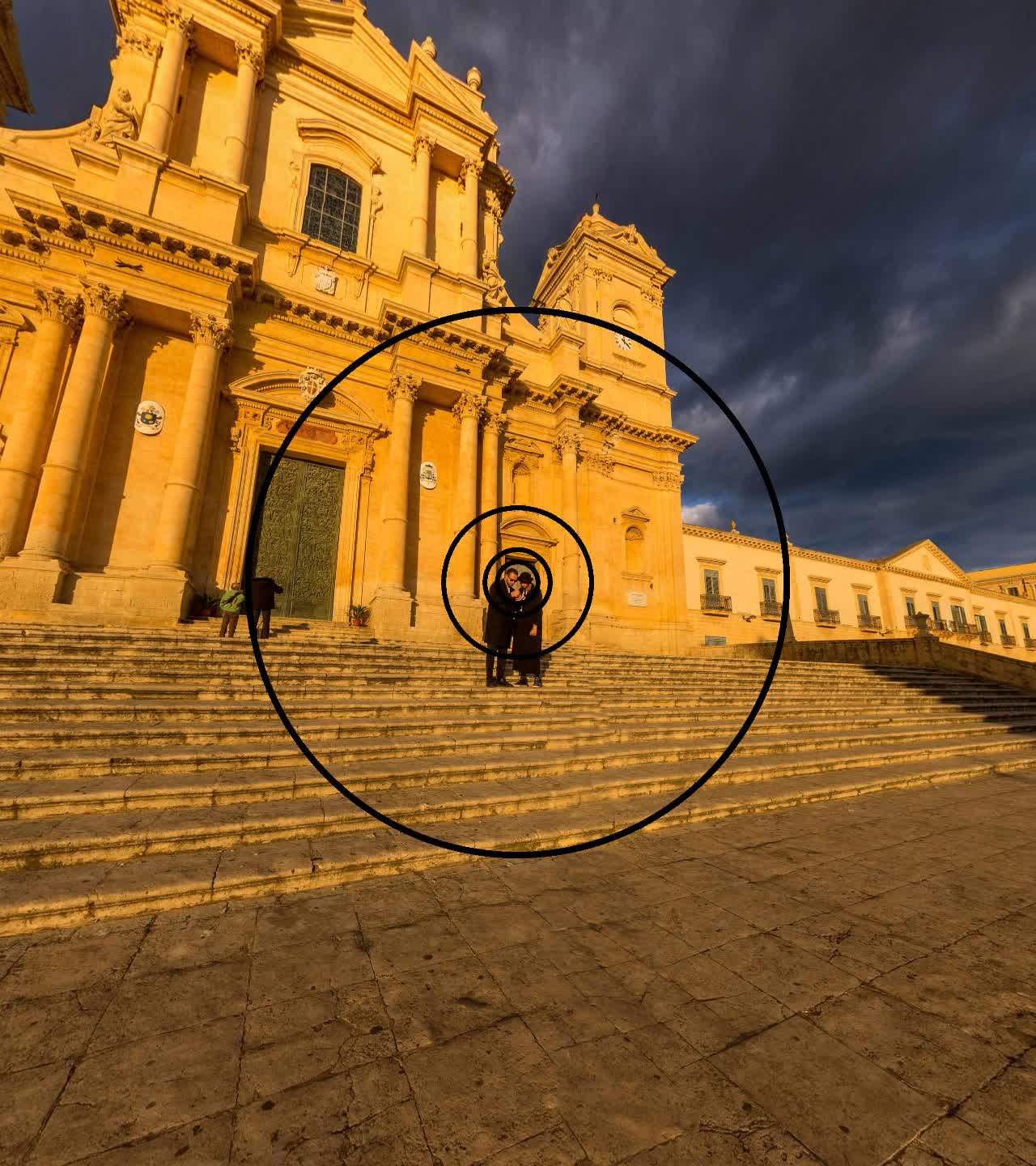}}
		\caption{Boundaries of zones in viewports used in our experiment}
		\label{figViewport}
	\end{figure*}

	\section{Experiment Description}\label{SectExperiment}
	
	For the experiment, we used eight omnidirectional images, denoted by \textit{I1}$\sim$ \textit{I8}, as shown in Fig.~\ref{figImage}. Two images \textit{I5} and \textit{I7} were obtained on Flickr under of the Creative Commons (CC) copyrights. The other six images were selected from the SUN 360 Database~\cite{SUNDB1,SUNDB2}. The characteristics of these images are described in Table~\ref{tabFeatureSourImage}. It can be seen that the selected images cover various categories of capturing environment and presence of human. All these images were down sampled to the resolution of 8192$\times$4096. We asked 10 participants to freely observe the source images and then point out attractive objects. Based on the obtained results, we selected a foveation point corresponding to a viewport for each image.

	In order to generate stimuli of non-uniform quality, each image was first spatially divided into five zones, denoted $Z_1, Z_2, Z_3, Z_4,$ and $Z_5$. In particular, each zone represents an eccentricity interval as shown in Table~\ref{tabEcenIntervalZones}. It can be seen that zones $Z_1, Z_2, Z_3, Z_4,$ and $Z_5$ respectively correspond to the fovea, parafovea, perifovea, near periphery, and far periphery in the retina. Fig.~\ref{figViewport} illustrates the boundaries of the zones in the viewports used in our experiment.

	As described in Sect.~\ref{SectIntro}, we consider two basic scenarios of spatial quality changes. In the first scenario ($S\#1$), the center has higher quality than the periphery; and in the second scenario ($S\#2$), the center has lower quality than the periphery. 
	For each scenario, we used four quality variation patterns as shown in Table~\ref{tabSetZonesPattern}. In patterns \textit{P1}, \textit{P2}, \textit{P3}, and \textit{P4}, which belong to scenario $S\#1$, the number of high quality zones gradually increases from 1 to 4. In the remaining patterns (i.e., \textit{P5}, \textit{P6}, \textit{P7}, and \textit{P8}), which belong to scenario $S\#2$, the number of high quality zones gradually reduces from 4 to 1.

	In this study, we used one high quality level corresponding to the quality level of the source images, and four low quality levels corresponding to four blurring levels. These blurring levels were generated using Gaussian filters with a fixed filter size of 50 and four different standard deviations $\sigma$. For scenario $S\#1$, the four $\sigma$ values are 2, 4, 8, and 12. For scenarios $S\#2$, the four $\sigma$ values are 1, 2, 4, and 6. The difference between the two scenarios is due to the fact that blurring in zones close to the foveation point is easier to be perceived than in the others. The source and blurred images were then blended into stimuli of non-uniform quality.
	Specifically, the high quality zones in the stimuli consist of pixels of the source images, and the low quality zones are comprised of pixels of the blurred images. Similar to~\cite{ChihFoveated2017}, to prevent noticeable boundaries between low and high quality zones, belts with the width of 5~degrees between two adjacent zones having a quality switch were used as transition belts. The quality levels in these belts  smoothly change using a linear function. Totally, our database consists of 256 stimuli, which were rated in the below tests.

	To display the stimuli, we used a device set of a Samsung Galaxy S6 smartphone and a Samsung Gear VR headset  with the 96~degree field of view. The Samsung Galaxy S6 has the screen resolution of 2560$\times$1440 and the display size of 5.1 inches. For the Samsung Gear VR headset, the focal length of the lens is $F$=62mm, and the distances from the lens to the displayed viewports and the eyes are approximately $S_0$=25mm and $S_2$=10mm respectively.

	\begin{table}[t]
		\caption{ Quality Variation Patterns (HQ: High quality and LQ: Low quality)}
		\centering
		\resizebox{\columnwidth}{!}{%
			\begin{tabular}{|c|c|c|c|c|c|c|}
				\hline
				&  & \multicolumn{5}{c|}{\textbf{Quality levels of zones}} \\ \cline{3-7} 
				\multirow{-2}{*}{\textbf{\begin{tabular}[c]{@{}c@{}}Scenario\end{tabular}}} & \multirow{-2}{*}{\textbf{Pattern}} & \text{\text{\begin{tabular}[c]{@{}c@{}}$Z_1$\\ {[}0$^{\circ}$,2.5$^{\circ}$)\end{tabular}}} & \text{\text{\begin{tabular}[c]{@{}c@{}}$Z_2$\\ {[}2.5$^{\circ}$,4$^{\circ}$)\end{tabular}}} & \text{\text{\begin{tabular}[c]{@{}c@{}}$Z_3$\\ {[}4$^{\circ}$,9$^{\circ}$)\end{tabular}}} & \text{\text{\begin{tabular}[c]{@{}c@{}}$Z_4$\\ {[}9$^{\circ}$,30$^{\circ}$)\end{tabular}}} & \text{\text{\begin{tabular}[c]{@{}c@{}}$Z_5$\\ {[}30$^{\circ}$,$+\infty$)\end{tabular}}} \\ \hline
				& \textit{\textbf{P1}} & \cellcolor[HTML]{DAE8FC}HQ & \cellcolor[HTML]{FFFFFF}LQ & \cellcolor[HTML]{FFFFFF}LQ & \cellcolor[HTML]{FFFFFF}LQ & \cellcolor[HTML]{FFFFFF}LQ \\ \cline{2-7} 
				& \textit{\textbf{P2}} & \cellcolor[HTML]{DAE8FC}HQ & \cellcolor[HTML]{DAE8FC}HQ & \cellcolor[HTML]{FFFFFF}LQ & \cellcolor[HTML]{FFFFFF}LQ & \cellcolor[HTML]{FFFFFF}LQ \\ \cline{2-7} 
				& \textit{\textbf{P3}} & \cellcolor[HTML]{DAE8FC}HQ & \cellcolor[HTML]{DAE8FC}HQ & \cellcolor[HTML]{DAE8FC}HQ & \cellcolor[HTML]{FFFFFF}LQ & \cellcolor[HTML]{FFFFFF}LQ \\ \cline{2-7} 
				\multirow{-4}{*}{\textbf{S\#1}} & \textit{\textbf{P4}} & \cellcolor[HTML]{DAE8FC}HQ & \cellcolor[HTML]{DAE8FC}HQ & \cellcolor[HTML]{DAE8FC}HQ & \cellcolor[HTML]{DAE8FC}HQ & \cellcolor[HTML]{FFFFFF}LQ \\ \hline
				& \textit{\textbf{P5}} & \cellcolor[HTML]{FFFFFF}LQ & \cellcolor[HTML]{DAE8FC}HQ & \cellcolor[HTML]{DAE8FC}HQ & \cellcolor[HTML]{DAE8FC}HQ & \cellcolor[HTML]{DAE8FC}HQ \\ \cline{2-7} 
				& \textit{\textbf{P6}} & \cellcolor[HTML]{FFFFFF}LQ & \cellcolor[HTML]{FFFFFF}LQ & \cellcolor[HTML]{DAE8FC}HQ & \cellcolor[HTML]{DAE8FC}HQ & \cellcolor[HTML]{DAE8FC}HQ \\ \cline{2-7} 
				& \textit{\textbf{P7}} & \cellcolor[HTML]{FFFFFF}LQ & \cellcolor[HTML]{FFFFFF}LQ & \cellcolor[HTML]{FFFFFF}LQ & \cellcolor[HTML]{DAE8FC}HQ & \cellcolor[HTML]{DAE8FC}HQ \\ \cline{2-7} 
				\multirow{-4}{*}{\textbf{S\#2}} & \textit{\textbf{P8}} & \cellcolor[HTML]{FFFFFF}LQ & \cellcolor[HTML]{FFFFFF}LQ & \cellcolor[HTML]{FFFFFF}LQ & \cellcolor[HTML]{FFFFFF}LQ & \cellcolor[HTML]{DAE8FC}HQ \\ \hline
			\end{tabular}%
		}
		\label{tabSetZonesPattern}
	\end{table}

	In the tests, we used the Absolute Category Rating method~\cite{ITUTP913}, which is shown the best method in~\cite{ChihFoveated2017}. Before doing actual tests, participants were trained to get accustomed to the devices and the rating procedure. In addition, they were instructed to appropriately adjust devices to obtain the best experience. During the test process, the stimuli were randomly displayed one at a time. Note that, for a stimulus, the corresponding viewport displayed on HMD was fixed during the test. Participants were asked to look straight ahead at each viewport displayed directly in front of them to keep focusing on the center, where has an attractive object such as a human face or a flower vase. After stabilizing the gaze direction, each participant verbally gave a score with the grade scale from 1 (bad) to 5 (excellent) which was recorded by an assistant. 
	
	For each stimulus, the viewing duration was decided by the participants themselves to obtain more reliable rating scores. Commonly, the participants spent about 5~seconds for rating a stimulus and then took a break of 5~seconds. To avoid the negative impacts of fatigue and boredom, the tests were divided into 6 sessions conducted in different weeks. Each participant took part in only two sessions. The duration of each session was no more than 10 minutes. There were totally 62 participants between the ages of 20 and 30. A screening analysis of the obtained results was performed following Recommendation ITU-T P.913~\cite{ITUTP913}, and two participants were rejected. After discarding the scores of these two participants, each stimulus was scored by 20 valid participants. The mean opinion score (MOS) of a stimulus is the average score of the valid participants.

	\begin{figure}[t]
		\centering
		\includegraphics[width=0.8\columnwidth]{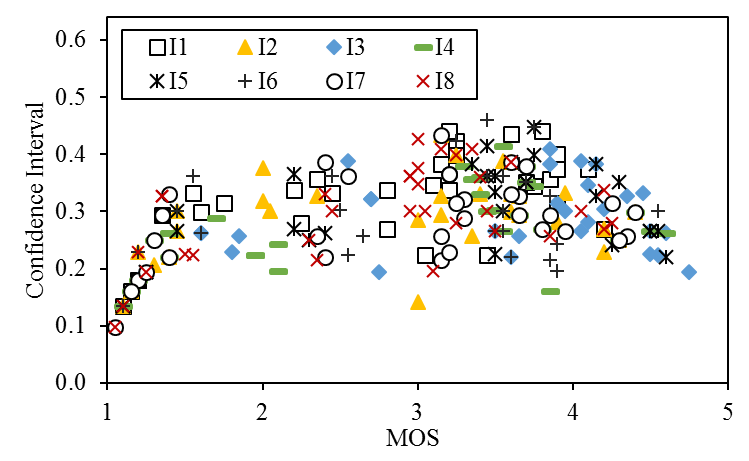}%
		\caption{95\% confidence intervals of MOS values}
		\label{figCI}
	\end{figure}

	The 95\% confidence intervals of the MOS values are shown in Fig.~\ref{figCI}. We can see that the scores cover fully the value range from 1 to nearly 5. Generally, the confidence intervals are  smaller at the two ends of the grade scale. This is because the participants are more confident in rating stimuli of very high (or low) quality.

	\section{Analysis of Perceptual Behaviors in Zones}\label{SectAnalysis}
	\subsection{Quantifying impacts of zones}\label{subSectZWQM}
	In this part, we present a zone-weighted formulation which will be used to analyze the impacts of different zones  on the perceptual quality of omnidirectional images. 
	In general, the virtual viewport is divided into $K$ zones $\{Z_{k}|1\le k\le K\}$, each consists of $N_k$ pixels with the corresponding eccentricities $e \in [e_{k-1},e_{k})$. Currently, we use $K=5$ as described in Sect.~\ref{SectExperiment}. Each zone $Z_k$ is then assigned a weight $\{w_{k}|1\le k\le K\}$ representing the impact of that zone on human perception of quality. Note that $\sum_{k=1}^{K}w_k=1$.

	Let $V(x_M,y_M)$ and $G(x_M,y_M)$ respectively be the values of pixel $M=(x_M,y_M)$ in the displayed viewports of the original and distorted images. The values of the corresponding pixel $M'=(x_{M'},y_{M'})$ in the virtual viewports of the original and distorted images are respectively calculated by the following equations. 
	\begin{equation}
	V'(x_{M'},y_{M'}) = V(x_M,y_M).
	\end{equation}
	\begin{equation}
	G'(x_{M'},y_{M'}) = G(x_M,y_M).
	\end{equation} 
	
	The mean squared error (MSE) of pixels in zone $Z_k$ is computed by 
	\begin{equation}\resizebox{\hsize}{!}{$
		\textit{MSE}_{k} = \frac{\displaystyle{\sum_{x_{M'}=1}^{W_p'}}\sum_{y_{M'}=1}^{H_p'}[V'(x_{M'},y_{M'})-G'(x_{M'},y_{M'})]^2\times R_k(x_{M'},y_{M'})}{\displaystyle{\sum_{x_{M'}=1}^{W_p'}}\sum_{y_{M'}=1}^{H_p'}R_k(x_{M'},y_{M'})},$
	}
	\end{equation}
	where
	\begin{equation}
	R_k(x_{M'},y_{M'}) = \left\{\begin{matrix}
	1,& \text{if}~~~e_{k-1} \le e(x_{M'},y_{M'}) < e_k\\ 
	0,& \text{otherwise}
	\end{matrix}\right.
	\end{equation}

	The zone-weighted formulation, called \wpsnr, is given by
	\begin{equation}
	\textit{\wpsnr} = 10 \log_{10} \left(\dfrac{\textit{MAX}^2}{\sum_{k=1}^{K}(w_k \times \textit{MSE}_k)}\right)~~~~~~~~\tagaddtext{[dB],}    
	\end{equation} 
	where \textit{MAX} is the maximum possible pixel value. Here we set \textit{MAX} to 255 as the bit depth of pixels is 8 bits in our experiment. 
	
	In some previous studies \cite{Huyen_VRMetrics,NonRegression}, it was shown that four-parameter and five-parameter logistic functions are good mappings between objective quality metrics and MOS. In this work, we deployed the following five-parameter logistic function to map the \wpsnr values and the MOS values in our database.
	\begin{equation}\label{Eq5param}
	y = \beta_1 \left( \frac{1}{2} - \frac{1}{1+e^{\beta_2(x-\beta_3)}} \right) +\beta_4x+\beta_5,
	\end{equation}
	where $\{\beta_i|i\in\{1,2,...,5\}\}$ are parameters to be fitted.  
	The values of the parameters $\beta_i$'s and the weights ${w_{k}}$'s were determined by means of least squares fitting as in~\cite{QoE_YWAngQSTAR}.

	\subsection{Discussion}\label{SectDisWeight}
\begin{figure}[t]
		\centering
		\includegraphics[width=0.8\columnwidth]{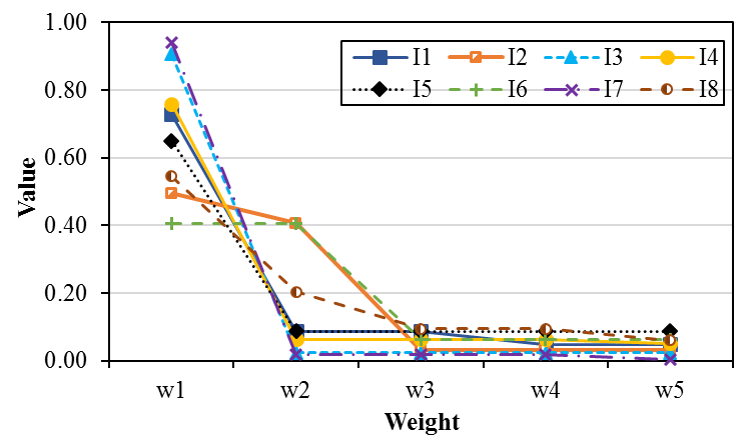}%
		\caption{Weights of zones for each source image}
		\label{figWeighteachI}
	\end{figure}

	To quantify the impact of each zone taking into account the effects of content characteristics, the weights ${w_k}$'s are derived for each source image by fitting using the above five-parameter logistic function  with the stimuli of that image only. The obtained values of the weights are shown in Fig.~\ref{figWeighteachI} and Table~\ref{tabWeighteachI}. The correlation coefficients including Pearson Correlation Coefficient (PCC) and Root Mean Square Error (RMSE), which are used to quantify the performance of the fitting between the \wpsnr formulation and the MOS, are shown in Table~\ref{tabPerZWQM}. We can see that, for all the source images, the PCC values are very high and the RMSE values are very low. In particular, the lowest PCC value is 0.97 while the highest RMSE value is 0.27. This means that the fitting to obtain the weights is reliable.

	From Table~\ref{tabWeighteachI}, it can be seen that, except $w_1$ and $w_2$, all the other weights are small (i.e., $\le 0.095$). That means the zones outside the eccentricity of 4 degrees have little impacts on the perceptual quality. Among the weights, $w_1$ is usually highest, which is consistent with the fact that the fovea region of the retina has the highest cone density. Also, because $w_1 \ge w_2 \ge w_3 \ge w_4 \ge w_5$, distortions closer to the center have more significant effects on the perceptual quality than distortions far from the center. 

	\begin{table}[t]
		\centering
		\caption{Weights of zones for each source image}
		\label{tabWeighteachI}
		\begin{tabular}{|c|c|c|c|c|c|}
			\hline
			\multirow{2}{*}{\textbf{Image}} & \multicolumn{5}{c|}{\textbf{Weight}}                                                                             \\ \cline{2-6} 
			& \textit{\textbf{w1}} & \textit{\textbf{w2}} & \textit{\textbf{w3}} & \textit{\textbf{w4}} & \textit{\textbf{w5}} \\ \hline
			\textit{\textbf{I1}}            & 0.728                & 0.088                & 0.088                & 0.048                & 0.048                \\ \hline
			\textit{\textbf{I2}}            & 0.495                & 0.407                & 0.033                & 0.033                & 0.032                \\ \hline
			\textit{\textbf{I3}}            & 0.905                & 0.024                & 0.024                & 0.024                & 0.024                \\ \hline
			\textit{\textbf{I4}}            & 0.759                & 0.063                & 0.063                & 0.063                & 0.052                \\ \hline
			\textit{\textbf{I5}}            & 0.650                & 0.087                & 0.087                & 0.087                & 0.087                \\ \hline
			\textit{\textbf{I6}}            & 0.404                & 0.404                & 0.064                & 0.064                & 0.064                \\ \hline
			\textit{\textbf{I7}}            & 0.941                & 0.019                & 0.019                & 0.019                & 0.003                \\ \hline
			\textit{\textbf{I8}}            & 0.545                & 0.204                & 0.095                & 0.095                & 0.061                \\ \hline
		\end{tabular}%
	\end{table}
	
	\begin{table}[t]
		\caption{Performance of fitting between the \wpsnr formulation and MOS}
		\label{tabPerZWQM}
		\centering
		\resizebox{0.45\textwidth}{!}{%
			\begin{tabular}{|l|l|l|l|l|l|l|l|l|}
				\hline
				\multicolumn{1}{|c|}{} & \multicolumn{1}{c|}{\textbf{I1}} & \multicolumn{1}{c|}{\textbf{I2}} & \multicolumn{1}{c|}{\textbf{I3}} & \multicolumn{1}{c|}{\textbf{I4}} & \multicolumn{1}{c|}{\textbf{I5}} & \multicolumn{1}{c|}{\textbf{I6}} & \multicolumn{1}{c|}{\textbf{I7}} & \multicolumn{1}{c|}{\textbf{I8}}\\ \hline
				\textbf{PCC} & 0.99&	0.99&	0.99&	0.97&	0.98&	0.99&	0.98&	0.99  \\ \hline
				\textbf{RMSE} & 0.15&	0.13&	0.14&	0.27&	0.24&	0.10&	0.20&	0.12  \\ \hline
			\end{tabular}%
		}
	\end{table}
	
	\begin{table*}
		\caption{Descriptions of objective quality metrics tested in this study. PW: Whether or not the metric differentiates pixels' contributions. FF: Whether or not the metric takes into account the foveation feature.}
		\label{tabDesOQMetrics}
		\resizebox{\textwidth}{!}{%
			\begin{tabular}{|m{1in}|M{0.2in}|M{0.2in}|m{5.2in}|}
				\hline
				\multicolumn{1}{|c|}{\textbf{Metrics}} & \textbf{PW}  & \textbf{FF}  & \multicolumn{1}{c|}{\textbf{Description}} \\ \hline
				\textbf{\mseVP		}					&  No & No	& Mean Squared Error, Calculated based on visble pixels of a viewport with equal weights  \\ \hline
				\textbf{\psnrVP		}					&  No & No	& Viewport-PSNR, Calculated based on visble pixels of a viewport with equal weights \\ \hline
				\textbf{\ssimVP		}~\cite{ssim}		&  No & No	& Structural SIMilarity, Calculated based on the concept of structural similarity \\ \hline
				\textbf{\mssimVP	}~\cite{msssim}		&  No & No	& Multi-scale SSIM, Calculated based on similar measures computed at different resolutions (or multi-scales) of a viewport \\ \hline
				\textbf{\uqiVP		}~\cite{uqi}		&  No & No	& Universal Image Quality,  Modeling any distortion as a combination of three different factors including loss of correlation, luminance distortion, and contrast distortion\\ \hline
				\textbf{\vifpVP		}~\cite{vif}		&  No & No	& \multirow{2}{*}{ \parbox{5in}{Visual Information Fidelity in the pixel domain (\textit{VIFp}) and the wavelet domain (\textit{VIF}), Calculated based on the connections between image information and visual quality}}\\  &&& \\ \cline{1-3}
				\textbf{\vipVP}~\cite{vif}				&  No & No	& \\ \hline
				\textbf{\nqmVP		}~\cite{nqm}		&  No & No 	& Noise Quality Measure, Signal-to-Noise Ratio of the restored distorted image with respect to the model restored image\\ \hline
				\textbf{\iwPnsrVP	}~\cite{iwSsim}		& Yes & No	& Information content Weighted PSNR, Combining information content weighting with PSNR measures  \\ \hline
				\textbf{\iwSsimVP	}~\cite{iwSsim}		& Yes & No	& Information content Weighted SSIM, Combining information content weighting with MS-SSIM measures \\ \hline
				\textbf{\fsimVP		}~\cite{fsim}		& Yes & No	& Feature similarity, Combining low-level feature weighting with local similarity measures \\ \hline
				\textbf{\fsimCVP	}~\cite{fsim}		& Yes & No	& Feature similarity incorporating the chromatic information, Combining low-level feature weighting with local similarity measures\\ \hline
				\textbf{\rfsimVP	}~\cite{rfsim}		& Yes & No	& Riesz Transforms based Feature Similarity, Combining low-level feature weighting based on Riesz Transforms with local similarity measures \\ \hline
				\textbf{\srSimVP	}~\cite{srsim}		& Yes & No	& Spectral Residual based Similarity, Calculated based on a  spectral residual visual saliency model\\ \hline
				\textbf{\fwqiVP		}~\cite{fwqi}		& Yes & Yes 	& Foveated Wavelet image Quality Index (FWQI), Calculated based on wavelet coefficients in the discrete wavelet transform domain using the foveation-based error sensitivity model as a weighting function.\\ \hline
				\textbf{\wsnrVP		}~\cite{nqm}		& Yes & Yes 	& Weighted Signal-to-Noise Ratio, the ratio of the average weighted signal power to the average weighted noise power, where the weighting function is the contrast sensitivity function \\ \hline
				\textbf{\fwSnrVP	}~\cite{LeeFWSNR}	& Yes & Yes 	& Foveal Weighted Signal-to-Noise Ratio, Combining weighting for each pixel by the local frequency at that pixel with WSNR measures \\ \hline
				\textbf{\fpsnrVP	}~\cite{LeeFPSNR}	& Yes & Yes 	& Foveal Peak Signal-to-Noise Ratio, Combining weighting for each pixel by the local frequency at that pixel with PSNR measures \\ \hline
				\textbf{\fSsimVP	}~\cite{fssim}		& Yes & Yes 	& Foveal-SSIM, Combining weighting for each macroblock based on the local frequency of pixels in that macroblock with SSIM measures\\ \hline
			\end{tabular}%
		}
	\end{table*}
	

	\begin{figure*}[!t]
		\centering
		\subfloat[PCC]{\includegraphics[width=0.48\textwidth]{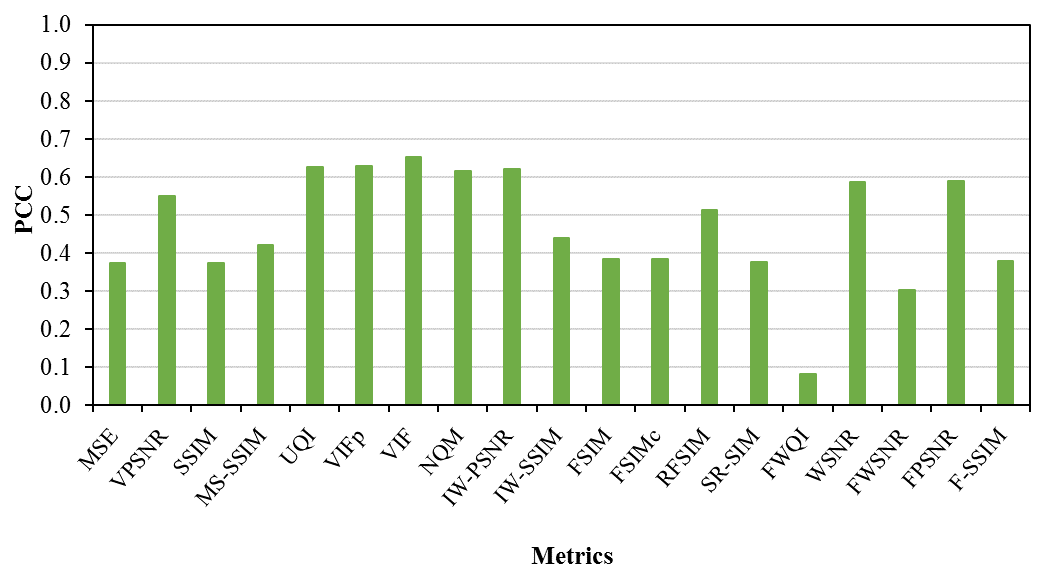}}
		\hfil
		\subfloat[RMSE]{\includegraphics[width=0.48\textwidth]{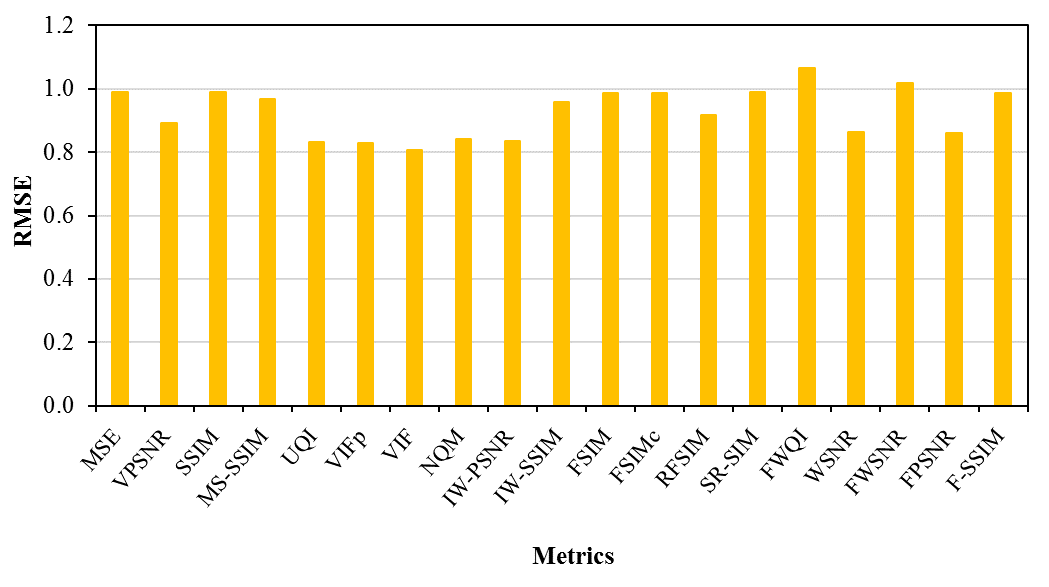}}
		\caption{Performances of objective quality metrics}
		\label{figPerOQMetric}
	\end{figure*}
	
	\begin{table*}
		\caption{Performances of metrics calculated with the stimuli of each source image. The bold numbers show the metrics having the highest performance for each source image.}
		\label{tabPerOQMetric}
		\centering
		\resizebox{\textwidth}{!}{%
			\begin{tabular}{|m{0.7in}|c|c|c|c|c|c|c|c|c|c|c|c|c|c|c|c|}
				\hline
				\multirow{2}{*}{\textbf{Metrics}} & \multicolumn{8}{c|}{\textbf{PCC}} & \multicolumn{8}{c|}{\textbf{RMSE}} \\\cline{2-17}  
				& \textit{\textbf{I1}} & \textit{\textbf{I2}} & \textit{\textbf{I3}} & \textit{\textbf{I4}} & \textit{\textbf{I5}} & \textit{\textbf{I6}} & \textit{\textbf{I7}} & \textit{\textbf{I8}}	& \textit{\textbf{I1}} & \textit{\textbf{I2}} & \textit{\textbf{I3}} & \textit{\textbf{I4}} & \textit{\textbf{I5}} & \textit{\textbf{I6}} & \textit{\textbf{I7}} & \textit{\textbf{I8}}  \\ \hline
				\textbf{\mseVP		}& 0.68 & 0.64 & 0.76 & 0.64 & 0.74 & 0.57 & 0.60 & 0.70 &                                                                  																 0.70 & 0.82 & 0.75 & 0.86 & 0.80 & 0.80 & 0.83 & 0.68 \\ \hline
				\textbf{\psnrVP		}& 0.69 & 0.56 & 0.69 & 0.68 & 0.80 & 0.69 & 0.57 & 0.60 &                                                              																	 0.69 & 0.88 & 0.83 & 0.81 & 0.70 & 0.70 & 0.85 & 0.76 \\ \hline
				\textbf{\ssimVP		}& 0.42 & 0.42 & 0.48 & 0.43 & 0.53 & 0.41 & 0.38 & 0.38 &                                                              																	 0.87 & 0.96 & 1.01 & 1.02 & 1.00 & 0.89 & 0.96 & 0.88 \\ \hline
				\textbf{\mssimVP	}& 0.47 & 0.48 & 0.52 & 0.49 & 0.60 & 0.45 & 0.43 & 0.45 &                                                              																	 0.84 & 0.93 & 0.99 & 0.99 & 0.95 & 0.87 & 0.94 & 0.84 \\ \hline
				\textbf{\uqiVP		}& 0.59 & 0.64 & 0.72 & 0.75 & 0.68 & 0.68 & 0.54 & 0.76 &                                                              																	 0.77 & 0.81 & 0.80 & 0.73 & 0.87 & 0.72 & 0.87 & 0.62 \\ \hline
				\textbf{\vifpVP		}& 0.63 & 0.67 & \textbf{0.79} & 0.75 & 0.70 & 0.62 & 0.61 & 0.61 &                                                              															 0.75 & 0.79 & \textbf{0.70} & 0.73 & 0.85 & 0.77 & 0.82 & 0.75 \\ \hline
				\textbf{\vipVP		}& 0.63	& 0.68 & 0.71 & 0.66 & 0.70	& 0.62 & 0.64 & 0.79 &																																	 0.74 & 0.77 & 0.81 & 0.83 & 0.84 & 0.76 & 0.80 & 0.58 \\ \hline
				\textbf{\nqmVP		}& 0.62 & 0.74 & 0.73 & {0.69} & {0.69}	& 0.66 & \textbf{0.69} & 0.69 &																														 0.75 & 0.71 & 0.79 & 0.81 & 0.85 &	0.73 & \textbf{0.75} & 0.69 \\ \hline
				\textbf{\iwPnsrVP	}& 0.63 & 0.53 & 0.53 & 0.64 & 0.60 & 0.67 & 0.65 & 0.65 &                                                              																	 0.74 & 0.90 & 0.98 & 0.85 & 0.95 & 0.72 & 0.79 & 0.72 \\ \hline
				\textbf{\iwSsimVP	}& 0.45 & 0.46 & 0.49 & 0.48 & 0.55 & 0.45 & 0.43 & 0.43 &                                                              																	 0.85 & 0.94 & 1.01 & 0.97 & 0.99 & 0.87 & 0.94 & 0.85 \\ \hline
				\textbf{\fsimVP		}& 0.40 & 0.41 & 0.48 & 0.43 & 0.48 & 0.42 & 0.39 & 0.38 &                                                              																	 0.88 & 0.96 & 1.02 & 1.01 & 1.04 & 0.88 & 0.96 & 0.87 \\ \hline
				\textbf{\fsimCVP	}& 0.40 & 0.41 & 0.48 & 0.43 & 0.48 & 0.42 & 0.38 & 0.38 &                                                              																	 0.88 & 0.96 & 1.02 & 1.01 & 1.04 & 0.88 & 0.96 & 0.87 \\ \hline
				\textbf{\rfsimVP	}& 0.59 & 0.53 & 0.55 & 0.51 & 0.62 & 0.51 & 0.50 & 0.52 &                                                              																	 0.78 & 0.90 & 0.97 & 0.96 & 0.93 & 0.84 & 0.90 & 0.81 \\ \hline
				\textbf{\srSimVP	}& 0.36 & 0.41 & 0.43 & 0.42 & 0.45 & 0.42 & 0.35 & 0.39 &	                                                            																	 0.89 & 0.97 & 1.04 & 1.00 & 1.07 & 0.89 & 0.98 & 0.87 \\ \hline
				\textbf{\fwqiVP		}&  {{0.10}}& {{0.04}}& {{0.01}}& {{0.09}}& {{0.11}}& {{0.27}}& {{0.31}}& {{0.13}}& 				 																						 {{0.95}}& {{1.06}}& {{1.16}}& {{1.10}}& {{1.18}}& {{0.94}}& {{1.01}}& {{0.94}}	 \\ \hline
				\textbf{\wsnrVP		}& 0.66&	0.66&	0.53&	0.61&	\textbf{0.83}&	0.67&	0.51&	0.65&																														 0.72&	0.80&	0.98&	0.87&	\textbf{0.65}&	0.72&	0.90&	0.72 \\ \hline
				\textbf{\fwSnrVP	}& 0.67 & 0.62 & 0.63 & 0.59 & 0.42 & 0.65 & 0.62 & 0.58 &                                                              																	 0.71 & 0.83 & 0.90 & 0.90 & 1.07 & 0.75 & 0.82 & 0.77 \\ \hline
				\textbf{\fpsnrVP	}& \textbf{0.84} & \textbf{0.86} & 0.59 & \textbf{0.84} & 0.81 & \textbf{0.99} & 0.59 & \textbf{0.98} &                 																	 {\textbf{0.52}} & {\textbf{0.54}} & 0.93 & {\textbf{0.60}} & 0.70 & {\textbf{0.12}} & 0.84 & {\textbf{0.20}} \\ \hline
				\textbf{\fSsimVP	}& 0.41 & 0.42 & 0.44 & 0.41 & 0.51 & 0.42 & 0.40 & 0.38 &                                                              																	 0.87 & 0.96 & 1.04 & 1.01 & 1.01 & 0.89 & 0.95 & 0.87 \\ \hline
			\end{tabular}%
		}
	\end{table*}

	\begin{figure*}
		\centering\centering
		\subfloat[\fwqiVP]{\includegraphics[width=0.3\textwidth]{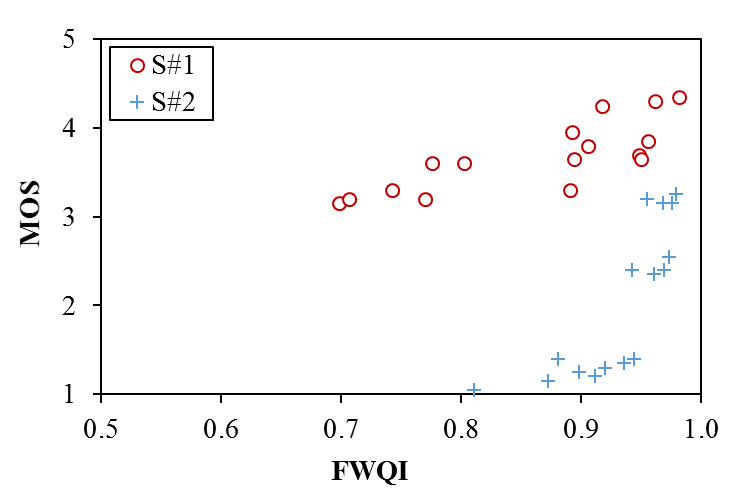}}
		\hfil
		\subfloat[\wsnrVP]{\includegraphics[width=0.3\textwidth]{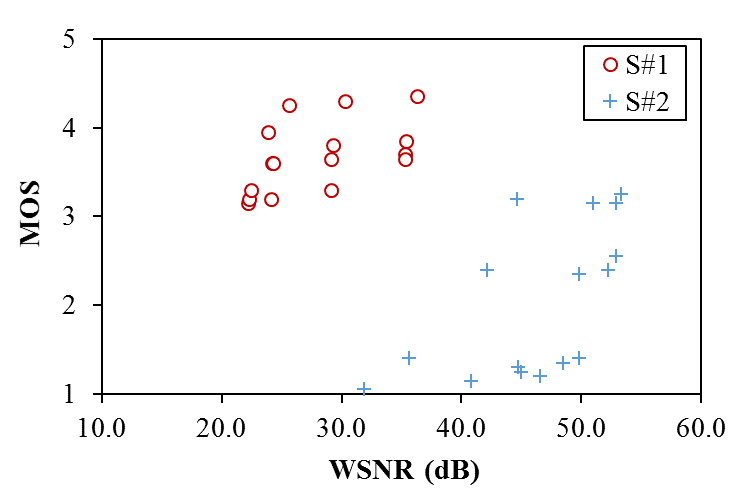}}
		\hfil
		\subfloat[\fwSnrVP]{\includegraphics[width=0.3\textwidth]{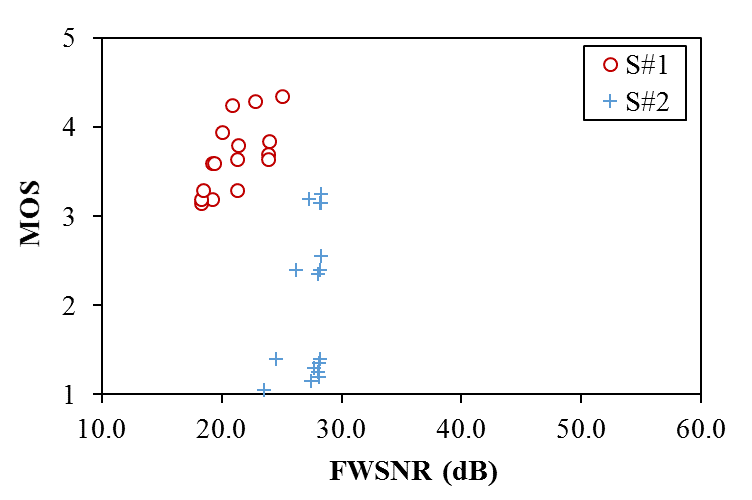}}
		\hfil
		\subfloat[\fpsnrVP]{\includegraphics[width=0.3\textwidth]{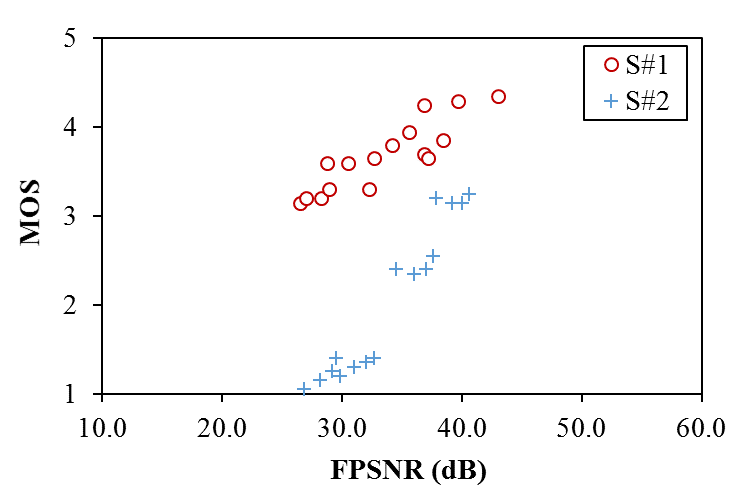}\label{FPSNRI7}}
		\hfil
		\subfloat[\fSsimVP]{\includegraphics[width=0.3\textwidth]{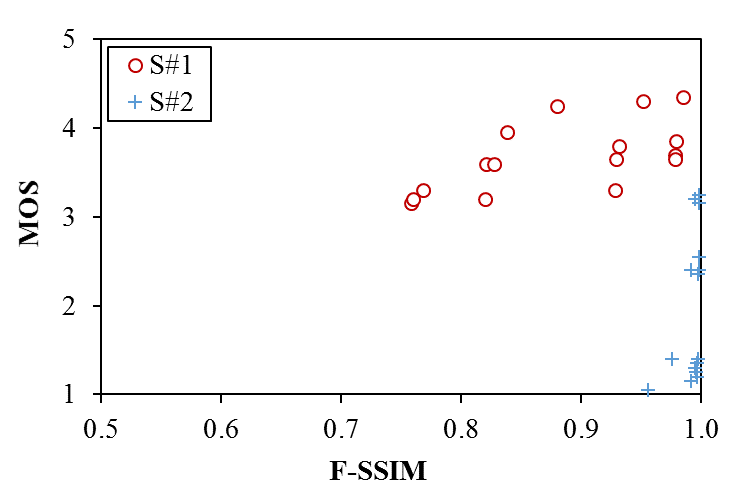}}
		\hfil
		\caption{Scatter plots of the values of the foveal quality metrics versus the MOS values   for  image \textit{I7}}
		\label{figScatter}
	\end{figure*}

	Based on Fig.~\ref{figWeighteachI}, it is interesting that the value of $w_1$ actually varies in a wide range. Also, with some images, the value of $w_2$ is insignificant. Usually, the higher the value of $w_1$ is, the lower the value of $w_2$ becomes.
	More specifically, with images \textit{I3} and \textit{I7}, the values of $w_1$ are very high. This may be because the participants focus primarily on the small face at the center of the viewports. Such phenomenon was also observed in~\cite{Lausane2011FoveatedCodingEvaluation}. In particular, it was found that a talking face is strongly attractive to human attention~\cite{Lausane2011FoveatedCodingEvaluation}. In addition, in these viewports, there are no other interesting objects near the center. 
	With images \textit{I1} and \textit{I4}, the participants may also pay some attention to other objects near the center (e.g., another face in image \textit{I1}), so the values of $w_1$ are lower than those of images \textit{I3} and \textit{I7}. With images \textit{I5} and \textit{I8}, the center's object is not very clear (small faces in image \textit{I8}) or not very attractive (a house in image \textit{I5}), resulting in lower values of $w_1$.
	Especially, with images \textit{I2} and \textit{I6}, the values of $w_2$ are comparable to those of $w_1$. In these images, the participants may look at a large central area rather than zone $Z_1$ only. The reason is that, in image \textit{I2}, the clock at the center is larger than zone $Z_1$; and in image \textit{I6}, the object at the center does not stand out from the neighboring area.

	From the above, we can see that the perceptual quality is affected by two key factors. The first is the sensitivity of human eyes. Especially, in the considered context, zones $Z_1$ and $Z_2$ are much more important than the other zones. The second is content characteristics. In particular, the values of $w_1$ and $w_2$ vary widely according to 1) the attractiveness and 2) the size of the central object, as well as 3) the presence of neighboring objects.

	\section{Evaluation of Quality Metrics}\label{SectEvaluation}
	In this part, by using our database, we evaluate the performances of nineteen existing objective quality metrics (OQM). \textcolor{black}{The goal is to examine whether existing metrics, especially foveal quality metrics, are effective for quality assessments of omnidirectional images with non-uniform quality.}

	\subsection{Description of metrics}
	Table~\ref{tabDesOQMetrics} shows the notations and descriptions of the nineteen metrics considered in this study. In this table, the PW column indicates whether a metric differentiates the contributions of different pixels; and the FF column indicates whether a metric takes into account the foveation feature of the human eye. Because the implementations of the \fwqiVP, \fwSnrVP, \fpsnrVP, and \fSsimVP metrics are not publicly available, we implemented them based on the corresponding publications~\cite{fwqi,LeeFWSNR,LeeFPSNR,fssim}. For the remaining metrics, we used the implementations provided by the original authors. 
	
	It is worth noting that all of these metrics were proposed to calculate for all pixels in a traditional image. In this study, these metrics were calculated for viewports only (i.e., visible pixels) of the omnidirectional images to reflect what is actually watched by viewers.  
	To extract the viewports, we used 360Lib software developed by Joint Video Experts Team (JVET)~\cite{360Lib}.  In addition, geometric parameters in these metrics were calculated based on the equations presented in Subsect.~\ref{subSectViewingGeometry}.

	In order to evaluate the performances of the OQM metrics, we used two performance metrics of  Pearson Correlation Coefficient (PCC) and Root Mean Square Error (RMSE). Similar to~\cite{NonRegression}, a nonlinear regression was applied to map the OQM values to the MOS values using the five-parameter logistic function (i.e., Equation (\ref{Eq5param})) mentioned in Subsect.~\ref{subSectZWQM}. 
	
	\subsection{Discussion}
	
	Fig.~\ref{figPerOQMetric} shows the PCC and RMSE values of the OQM metrics when fitting with all the MOSs in our database. It can be seen that all the metrics have very low PCC values (i.e., PCC $<0.70$) and very high RMSE values  (i.e., RMSE $>0.80$). Even the foveal quality metrics (namely \fwqiVP, \wsnrVP, \fwSnrVP, \fpsnrVP, and \fSsimVP) have bad PCC values (i.e., from 0.08 to 0.59). This means that the investigated metrics are not effective to assess the perceptual quality of omnidirectional images with non-uniform quality.

	Similar to the previous analysis related to the \wpsnr formulation, it is important to understand the performances of the metrics for each source image. Table~\ref{tabPerOQMetric} shows the performances of the metrics when fitting with the stimuli of each source image. It can be seen that, for all the metrics, the PCC and RMSE values are drastically variable across different source images. The bold numbers show the metrics having the highest performance for each source image. 
	
	Among the investigated metrics, the \fpsnrVP metric has the highest PCC values for five source images (i.e.,  \textit{I1}, \textit{I2}, \textit{I4}, \textit{I6}, and \textit{I8}). In addition, its PCC values for six images are quite good (i.e., PCC $>0.80$). Especially, with images \textit{I6} and \textit{I8}, the PCC values are nearly perfect (0.99 and 0.98). However, for two images \textit{I3} and \textit{I7}, its PCC values are very low (i.e., PCC $<0.70$), even lower than those of the \mseVP metric (i.e., 0.59 vs. 0.76 and 0.59 vs. 0.60), which is the simplest metric in practice. 
	
	As for the other quality metrics, their performances are mostly low. Even the other foveal quality metrics (i.e., except the \fpsnrVP metric) have lower performances than the non-foveal and simple metrics.
	
	To understand the actual behaviors of the foveal quality metrics that cause low performances, Fig.~\ref{figScatter} shows the scatter plots of the values of these metrics versus the MOS values for image \textit{I7}. 
	In this figure, we use different legends to differentiate the stimuli of scenario $S\#1$, where the center has higher quality, and the stimuli of scenario $S\#2$, where the center has lower quality. It is well-known that higher values of these metrics mean higher MOS values and better perceptual quality. From Fig.~\ref{figScatter}, we can see that the MOS values in scenario $S\#1$ are generally higher than those in scenario $S\#2$. However, for the \wsnrVP and \fSsimVP metrics, most of their values in scenario $S\#1$ are significant lower than those in scenario $S\#2$. For the remaining metrics, with the same MOS value, their corresponding values vary in a wide range. These result in the low performances of the foveal quality metrics.   
	
	
	From the above analysis, we can see that the investigated metrics are not effective to evaluate omnidirectional images of non-uniform quality. Though the \fpsnrVP metric (i.e the most feasible one) have very high performances in certain images, it performs even worse than the simple \mseVP metric in some other images.
	Moreover, the performances of all the quality metrics are not good across different images. This suggests that it is necessary to integrate content characteristics in these quality metrics.

	\section{Conclusions}\label{SectConclusion}
	
	In this paper, we have conducted subjective and objective quality assessments of omnidirectional images with non-uniform quality focusing on foveation feature of human eyes. Based on the obtained results and discussions, some findings can be summarized as follows.  
	\begin{itemize}
		\item The perceptual quality is affected by two key factors, which are the sensitivity of human eyes and content characteristics. 
		\item The zones of an image corresponding to the fovea and parafovea of human eyes are extremely important for the perceptual quality.
		\item Content characteristics including the attractiveness and the size of central object, as well as the presence of neighboring objects affect the quality perception.      
		\item The nineteen objective quality metrics considered in this study (including foveal quality metrics) are not effective to evaluate omnidirectional images with non-uniform quality.
		\item The performances of the investigated metrics vary drastically across different contents. 
	\end{itemize}
	
	For future work, further investigations with more content types and quality variation patterns will be conducted to derive better understanding of viewers' perceptual behaviors as well as the performances of existing metrics.

\end{document}